\documentclass[sigconf]{acmart}

\usepackage{booktabs} 

\usepackage{graphicx,epsfig,amsmath,amssymb,bm}
\usepackage{amssymb,balance}
\usepackage{amsmath}
\usepackage{algorithmicx}
\usepackage{algpseudocode}
\newtheorem{algorithm}{Algorithm}
\def\LSB{\left[}        
\def\RSB{\right]}       
\def\LB{\left(}         
\def\RB{\right)}        

\newfont{\bbb}{msbm10 scaled 500}

\newfont{\bb}{msbm10 scaled 1100}

\newcommand{\RR}{\mbox{\bb R}}

\newcommand{\av}{{\bf a}}
\newcommand{\bv}{{\bf b}}

\newcommand{\ev}{{\bf e}}

\newcommand{\iv}{{\bf i}}

\newcommand{\rv}{{\bf r}}

\newcommand{\uv}{{\bf u}}
\newcommand{\wv}{{\bf w}}
\newcommand{\vv}{{\bf v}}
\newcommand{\xv}{{\bf x}}
\newcommand{\yv}{{\bf y}}

\newcommand{\Am}{{\bf A}}
\newcommand{\Bm}{{\bf B}}
\newcommand{\Cm}{{\bf C}}

\newcommand{\Id}{{\bf I}}

\newcommand{\Km}{{\bf K}}

\newcommand{\Pm}{{\bf P}}
\newcommand{\Qm}{{\bf Q}}
\newcommand{\Rm}{{\bf R}}

\newcommand{\Wm}{{\bf W}}

\newcommand{\Xm}{{\bf X}}
\newcommand{\Ym}{{\bf Y}}

\newcommand{\deltav}{\hbox{\boldmath$\delta$}}

\newcommand{\defines}{{\,\,\stackrel{\scriptscriptstyle \bigtriangleup}{=}\,\,}}

\usepackage{bbm}
\usepackage{subcaption}
\usepackage{color}

\def\argmax{\operatornamewithlimits{arg\,max}}
\def\argmin{\operatornamewithlimits{arg\,min}}

\newcommand{\beqa}{\begin{eqnarray}}
\newcommand{\eeqa}{\end{eqnarray}}
\newcommand{\dsp}{\displaystyle}

\setcopyright{rightsretained}

\begin{document}
\title{Optimal Attack against Cyber-Physical Control Systems with Reactive Attack Mitigation}
\titlenote{This work was supported in part by the National Research
Foundation (NRF), Prime Minister's Office, Singapore, under
its National Cybersecurity R\&D Programme (Award
No. NRF2014NCR-NCR001-31) and administered by the
National Cybersecurity R\&D Directorate and in part by a
Start-up Grant at NTU.}

\copyrightyear{2017} 
\acmYear{2017} 
\setcopyright{acmcopyright}
\acmConference{e-Energy '17}{May 16-19, 2017}{Shatin, Hong Kong}\acmPrice{15.00}\acmDOI{http://dx.doi.org/10.1145/3077839.3077852}
\acmISBN{978-1-4503-5036-5/17/05}

\author{Subhash Lakshminarayana}
\affiliation{%
  \institution{Advanced Digital Sciences Center, Illinois at Singapore}
  \city{Singapore} 
  \postcode{138682}
}
\email{subhash.l@adsc.com.sg}

\author{Teo Zhan Teng}
\authornote{The work was conducted when Teo Zhan Teng was with the Advanced Digital Sciences Center, Illinois at Singapore.}
\affiliation{%
  \institution{GovTech Singapore}
  \city{Singapore} 
  \postcode{117438}
}
\email{teozt@hotmail.com}

\author{David K.Y. Yau}
\affiliation{%
  \institution{Singapore University of Technology and Design}
  \city{Singapore} 
  \postcode{487372}}
\email{david\_yau@sutd.edu.sg}

\author{Rui Tan}
\affiliation{
  \institution{Nanyang Technological University}
\city{Singapore} 
  \postcode{639798}}
\email{tanrui@ntu.edu.sg }

\begin{abstract}
This paper studies the performance and resilience of a cyber-physical control system (CPCS) with attack detection and reactive attack mitigation.
It addresses the problem of deriving an optimal sequence of false data injection attacks 
that maximizes the state estimation error of the system. The results provide basic understanding about the limit of the attack impact. The design of the optimal attack is based on a Markov decision process (MDP) formulation, which is solved efficiently using the value iteration method. Using the proposed framework, we quantify the effect of false positives and mis-detections on the system performance, which can help the joint design of the attack detection and mitigation. To demonstrate the use of the proposed framework in a real-world CPCS, we consider the voltage control system of power grids, and run extensive simulations using PowerWorld, a high-fidelity power system simulator, to validate our analysis. The results show that by carefully designing the attack sequence using our proposed approach, the attacker can cause a large deviation of the bus voltages from the desired setpoint. Further, the results verify the optimality of the derived attack sequence and show that, to cause maximum impact, the attacker must carefully craft his attack to strike a balance between the attack magnitude and stealthiness, due to the simultaneous presence of attack detection and mitigation.
\end{abstract}

\keywords{ Cyber-physical control system, Reactive attack mitigation, Resilience, Voltage control.}

\maketitle

\section{Introduction}
Critical infrastructures such as power grids and transportation systems
are witnessing growing adoption of modern information and communication technologies (ICTs) for autonomous operation. While these advancements have improved their operational efficiency, ICTs may also make them vulnerable to cyber attacks. Vulnerabilities in ICT systems were exploited in recent high-profile cybersecurity incidents such as the BlackEnergy \cite{Ukraine2016} and Dragonfly \cite{dragonfly2014} attacks against power grids  and the Stuxnet worm \cite{karnouskos2011} against nuclear plants. These attacks injected false sensor data and/or control commands to the industrial control systems and resulted in widespread damage to the physical infrastructures and service outages.  These incidents alert us to a general class of attacks called {\em false data injection} (FDI) against cyber-physical systems (CPS).

Attack detection and mitigation are two basic CPS security research problems, where the {\em attack detection} makes decisions in real time regarding the presence of an attack and {\em attack mitigation} isolates a detected attack and/or reduces its adverse impact on the system performance. 
CPSs often have various built-in anomaly detection methods that are effective in detecting simple fault-like FDI attacks, such as injecting surges, ramps, and random noises. However, critical CPSs (e.g., power grids)
are the target of sophisticated attackers (such as hostile national organizations), 
whose attacks are often well-crafted using detailed knowledge of the system and its anomaly detection methods. To avoid detection, the attacker can inject a sequence of attacks of small magnitude and gradually mislead the system to a sub-optimal and even unsafe state. 
However, due to the stochastic nature of the physical and measurement processes of CPSs, as well as the adoption of stringent, advanced attack detectors, the well-crafted attacks can be detected probabilistically \cite{Mo2015, vu2016}. Upon detecting an attack, mitigation should be activated to isolate the attack or maintain acceptable system performance in coexisting with the attack. 

Therefore, attack detection and mitigation are deeply coupled and they jointly define the system resilience against FDI attacks. On the one hand, a conservative detector may miss attacks, causing system performance degradation due to the mis-activation of attack mitigation. On the other hand, an aggressive detector may frequently raise false positives, triggering unnecessary mitigation actions in the absence of attacks, while attack mitigation generally needs to sacrifice the system performance to increase its robustness against attacks. Thus, it is important to understand the joint effect of attack detection and mitigation on the system performance, which serves as a basis for designing satisfactory detection-mitigation mechanisms. 
However, prior research on FDI attacks mostly study attack detection and mitigation separately \cite{LiuKwonHwang2012,Kwon2013,Mo2015,BaiGupta2014}, and falls short of capturing their joint effect on the system. The studies on attack detection \cite{LiuKwonHwang2012,Kwon2013,Mo2015} generally ignore the attack mitigation triggered by probabilistic detection of attacks, and its impact on the future system states. On the other hand, the studies on attack mitigation \cite{Barreto2013, Ma2013, ZhuBasar2015} assume that the attack has been detected, and ignore the probabilistic nature of the attack detection and any adverse impact of mis-activation or false activation of mitigation due to misdetections and false alarms.

As an early (but important) effort in closing the gap, we jointly consider attack detection and mitigation in the system defense. In particular, we study their joint effect from an attacker's perspective and investigate the largest system performance degradation that a sophisticated attacker can cause in the presence of such a detection-mitigation defense mechanism. Studying this largest performance degradation helps us quantify the limit of attack impact, and serves as an important basis for designing/comparing detection and mitigation strategies to protect critical infrastructures. 
However, the attacker faces a fundamental dilemma in designing his attack -- a large attack magnitude will result in high detection probability, thus nullifying the attack impact on the system (due to mitigation) whereas a small attack magnitude increases stealthiness but may do little damage. To achieve a significant impact, the attacker's injections must strike a balance between magnitude and stealthiness.

In this paper, we consider a general discrete-time linear time invariant (LTI) system with a feedback controller that computes its control decision based on the system state estimated by a Kalman filter (KF). For each time step, the controller uses a $\chi^2$ attack detector \cite{MEHRAChiSquare1971}, and activates mitigation actions upon detecting an attack. 
Following the Kerckhoffs's principle, we consider an attacker who accurately knows the system and its attack detection and mitigation methods. The attacker launches FDI attacks on the sensor measurements over an attack time horizon, aiming at misleading the controller into making erroneous control decisions. As the attack detection at each time step is probabilistic, we formulate the attacker's problem as a constrained stochastic optimization problem with an objective of maximizing the state estimation error over the attack time horizon, 
subject to a general constraint that the energy of the attack signal is upper-bounded. The solution to this problem naturally leads to an attack sequence that strikes a balance between attack magnitude and stealthiness to achieve the largest system performance degradation.

The main challenge in solving the aforementioned attacker's problem lies in the fact that the system state
at any time depends on all the past attack detection results, due to reactive attack mitigation. Thus, the optimal attack at any time must exhaustively account for all possible sequences of past detection results, which is computationally complex. Moreover, the probabilistic attack detection introduces additional randomness into the system dynamics. Our key observation to overcome these issues is that the system dynamics is Markovian and the attacker's injections at any time can be computed based on knowledge about it, which captures the impact of all the past detection results. To summarize, the main contributions of our work are as follows:
\begin{itemize}
\item We solve the aforementioned attacker's problem using a Markov decision process (MDP) framework. In our formulation, the sequential operations of probabilistic attack detection and mitigation are mapped to the MDP's state transition probabilities. The MDP is solved by state space discretization and using the \emph{value iteration} algorithm \cite{Puterman:1994}.

\item To illustrate our analysis, we use a real-world CPCS -- power grid voltage control -- as our case study. The voltage controller adjusts the pilot bus voltages to predefined setpoints based on voltage measurements by applying feedback control on the generators' reactive power outputs. 
In the presence of attack mitigation, the attacker injects false measurements into the system, aiming at deviating the pilot bus voltages. 
Extensive simulations using PowerWorld, a high-fidelity power simulator, show that the optimal attack sequence 
computed using our proposed approach causes the maximum deviation of the pilot bus voltages from the desired setpoint.

\item Based on the above framework, we also consider the problem of designing the detection threshold 
from the defender's perspective. To this end, we quantify the impact of false positives (FP) and misdetections (MD) via an extensive simulation-based study. Based on these costs, the attack detection threshold can be tuned to balance the performance downgrades due to FPs and MDs depending on the accuracy of the mitigation signal.

\end{itemize}

The remainder of the paper is organized as follows. Section~\ref{sec:Related} reviews related work. 
Section~\ref{sec:Sys_Model} describes the system model. Section~\ref{sec:Threat_Model} gives the problem formulation. 
Section~\ref{sec:Soln_Methods} describes the MDP-based solution methodology. 
Section~\ref{sec:MD_FP} analyzes the impact of FPs and MDs on the system performance. 
Section~\ref{sec:Sim_Res} presents the simulation results. Section~\ref{sec:Conclusion} concludes.

\begin{figure*}[!t]
\centering
\includegraphics[width=0.62\textwidth]{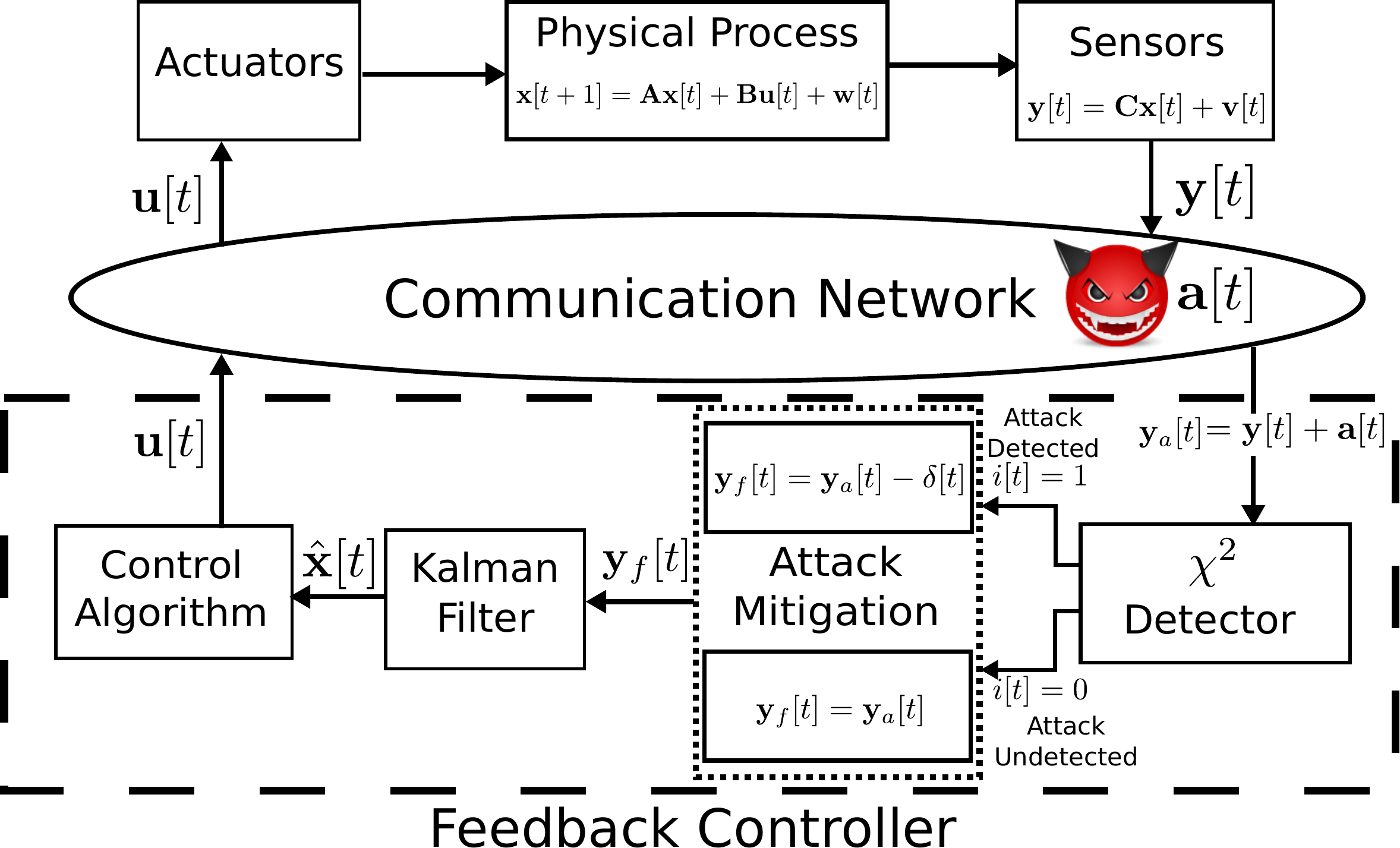}
\caption{Block diagram of the system model.}
\label{fig:sys_model}
\end{figure*}

\section{Related Work}
\label{sec:Related}
As mentioned earlier, most of the existing studies treat attack detection and mitigation problems separately. 
In the category of attack detection,
the performance degradation caused by stealthy attacks in a noiseless LTI system
has been analyzed \cite{Pasqualetti2013, FawziTAC2014}. Any deviation from the expected state trajectory in the deterministic system can be considered a fault or an attack. 
However, non-determinism and measurement noises experienced by real-world systems provide an opportunity for the attacker to masquerade his attack as natural noises, thereby rendering attack detection probabilistic. 
Research \cite{LiuKwonHwang2012}, \cite{Kwon2013}, and \cite{Mo2015} has studied 
the impact of stealthy false data injection (FDI) attacks against stochastic LTI systems, and derived optimal attack sequences that can cause the worst system performance degradation. Bai and Gupta \cite{BaiGupta2014} characterize a fundamental trade-off between the stealthiness level of an attack and the system performance degradation. However, these studies \cite{LiuKwonHwang2012,Kwon2013,Mo2015,BaiGupta2014} generally ignore the attack mitigation triggered by probabilistic detection of attacks and its impact on the future system states and attack detection.

In the category of attack mitigation, preventive and reactive mitigation strategies have been proposed \cite{Combita2015}. Preventive mitigation identifies vulnerabilities in the system design and removes them to prevent exploitation by attackers. For instance, in a power system,  
a set of sensors and their data links can be strategically selected and protected such that a bad data detection mechanism cannot be bypassed by FDI attacks against other sensors and their links that are not protected~\cite{Bobba2010, Dan2010}.
However, preventive mitigation provides static solutions only, which do not address the adaptability of strategic and knowledgeable attackers against critical infrastructures. Thus, in addition to preventative mitigation, it is important to develop reactive attack mitigation, i.e., countermeasures that are initiated after detecting an attack and tune the system based on the estimated attack activities. Reactive attack mitigation is mainly studied under game-theoretic settings \cite{Barreto2013, Ma2013}. Specifically, the attacker manipulates a set of sensor/control signals and aims at disrupting the system operation, while the defender responds by tuning the remaining system parameters to negate the attack or minimize its impact. However, most studies on reactive mitigation (e.g., \cite{Barreto2013, Ma2013, ZhuBasar2015}) assume that the attack has been detected, and ignore the impact of uncertain attack detection on the overall attack mitigation.
In contrast, our framework captures the interdependence between the 
attack detection and mitigation, and their joint impact on the system's dynamics and performance.

\section{Preliminaries} 
\label{sec:Sys_Model}
\subsection{System Model}

A block diagram of the system model is illustrated in Fig.~\ref{fig:sys_model}.
We consider a general discrete-time LTI system that evolves as
\begin{align}
\xv[t+1] &= \Am \xv[t] + \Bm \uv[t] + \wv[t], \label{eqn:process}
\end{align}
where $\xv[t] \in \RR^{n}$ is the system state vector, 
$\uv[t] \in \RR^{p}$ is the control input, and $\wv[t] \in \RR^{n}$ is the process noise at the $t$-th time slot. Matrices $\Am$ and $\Bm$ denote the propagation and control matrices, respectively.
The initial system state $\xv[0]$ and process noise $\wv[t]$ are independent Gaussian random variables. Specifically, $\xv[0] \sim \mathcal{N} (\bf{0},\Xm)$ and $\wv[t] \sim \mathcal{N} (\bf{0},\Qm),$ where $\mathbf{0} = [0, \ldots, 0]^T$ and $\mathbf{X}$ and $\mathbf{Q}$ are the covariance matrices.
The process described in \eqref{eqn:process} is observed through sensors deployed in the system, whose
observation at time $t$, denoted by $\yv[t] \in \RR^m$, is given by
\begin{align}
\yv[t] &= \Cm \xv[t] + \vv[t], \label{eqn:Obs}
\end{align}
where $\Cm \in \RR^{m \times n}$ is the measurement matrix and $\vv[t] \sim \mathcal{N} (\bf{0},\Rm)$ is the measurement noise at time $t$ and $\Rm$ is the covariance. We assume that $\vv[t]$ is
independent of $\xv[0]$ and $\wv[t].$
Moreover, we assume that the system in \eqref{eqn:process} is controllable and the measurement process in \eqref{eqn:Obs} is observable.

The controller uses a Kalman filter (KF) to estimate the system state based on the observations. The KF works as follows \cite{kailath2000linear}:
\begin{align}
\hat{\xv} [t+1] \!=\! \Am \hat{\xv}[t]\! +\! \Bm \uv[t] \! +\! \Km (\yv [t+1]\! -\! \Cm (\Am \hat{\xv}[t]\! +\! \Bm \uv[t])), \label{eqn:KF_est_nomit}
\end{align}
where $\hat{\xv}[t]$ is the estimate of the system state, $\Km$ denotes the steady-state Kalman gain given by $\Km =  \Pm_\infty \Cm^T(\Cm  \Pm_\infty\Cm^T + \Rm )^{-1}$, and 
the matrix $\Pm_\infty$ is the solution to the algebraic Riccati equation $\Pm_\infty = \Am\Pm_\infty\Am^T + \Qm - \Am\Pm_\infty\Cm^T(\Cm\Pm_\infty\Cm + \Rm)^{-1}\Cm\Pm_\infty\Am^T.$
We denote the KF estimation error at time $t$ by $\ev[t] = \xv[t] - \hat{\xv}[t].$

\subsubsection*{LTI Model in Power Systems}
The analysis in this paper is based on the general discrete-time LTI model described above. As a number of control loops found in a power system can be modeled using the LTI model, our analysis applies to these control loops.
In the following, we provide examples of a discrete-time LTI system, namely a power system's voltage control and generator 
swing equations. 

\begin{figure}[!t]
\centering
\includegraphics[width=0.45\textwidth]{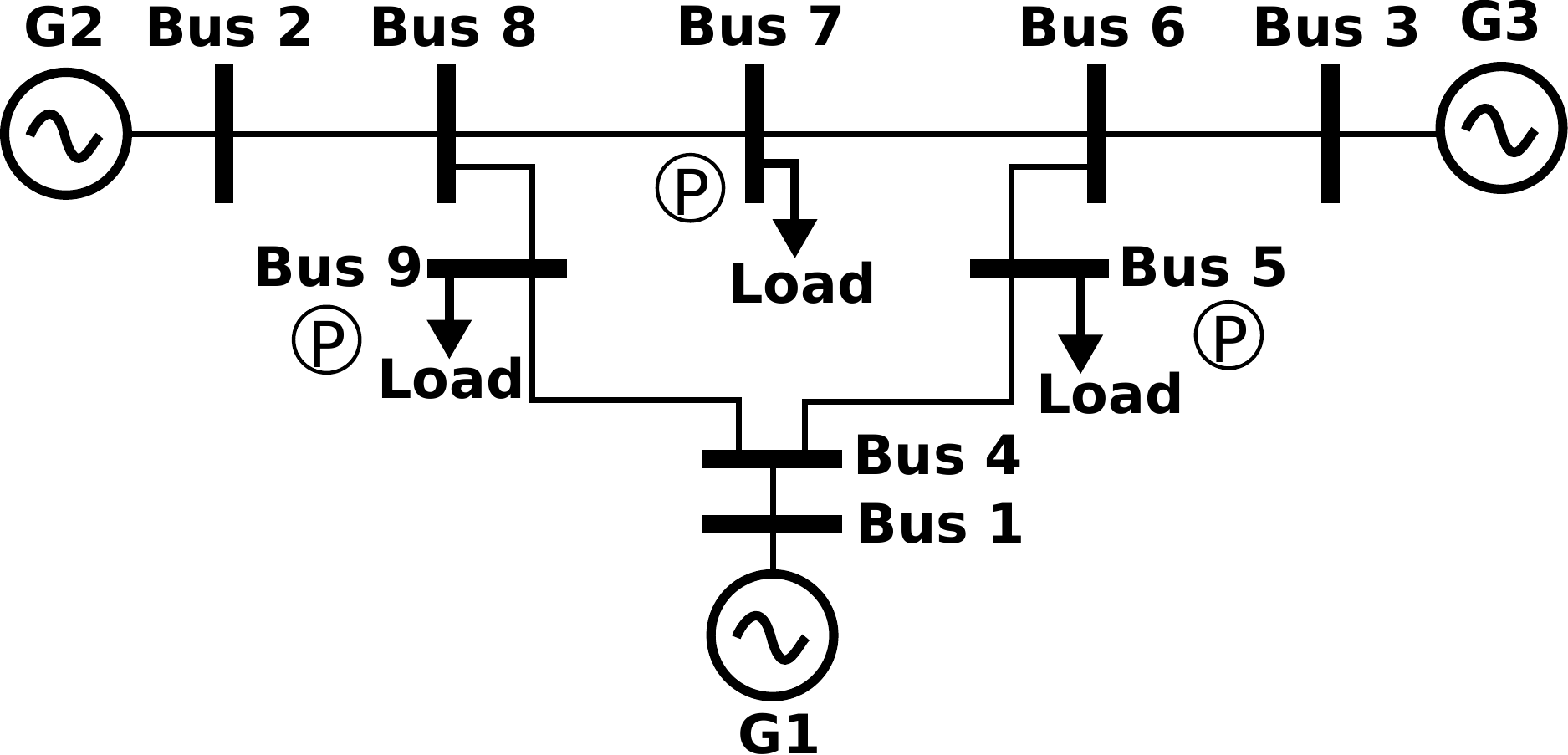}
\caption{IEEE 9-bus power system.}
\label{fig:9_bus}
\end{figure}

A power system consists of a set of buses (nodes) to which generators and loads are connected, and transmission lines that connect these buses. As an example, the IEEE $9$-bus test system is illustrated in Fig.~\ref{fig:9_bus}. 

\emph{Voltage control}: Power system voltage control refers to maintaining the voltages of selected critical buses (called {\em pilot buses} marked with ``P" in Fig.~\ref{fig:9_bus}) within safe operational limits by adjusting the 
output voltage of the generator buses \cite{Ilic1995}. It can be modeled as an LTI system described in Eqs. \eqref{eqn:process} and \eqref{eqn:Obs}. Specifically, the state vector $\xv[t]$ refers
to the voltages of the pilot buses at time $t,$ which should be maintained at a nominal
voltage denoted by $\xv_0.$
The control signal, which is applied at the generator buses, corresponds to the change in the generator bus voltages, i.e., $\uv[t] = \vv_G[t]-\vv_G[t-1],$ where $\vv_G[t]$ is a vector of the generator bus voltages. 
Under this model, the voltage control system can be approximated by an LTI system with $\Am = \Id$ \cite{Paul1987}, \cite{Ilic1995}. The control matrix $\Bm$ is 
an unknown parameter that can be estimated from real data traces (more details on estimating the matrix $\Bm$  will be presented in Section \ref{sec:Sim_Res}). 
Since the estimation cannot be perfect, the LTI model may be inaccurate, though the inaccuracies are small and can be captured as  process noise. 
Since the system state
can be directly measured by voltage sensors deployed at the pilot buses, the measurement matrix is an identity matrix, i.e., $\Cm = \Id.$
The system is bounded-input bounded-output stable if the control algorithm 
satisfies $\Bm \uv[t] = \alpha (\xv_0 - \xv[t])$ for $\alpha \in (0,1)$, and this control
is adopted in practical systems \cite{Paul1987}. However, as the sensor measurements are noisy, 
the controller cannot have perfect knowledge of the system state $\xv[t].$ Rather, the state is estimated
using the KF-based technique described in \eqref{eqn:KF_est_nomit}.
Based on the estimated state $\hat{\xv}[t],$ the control can be computed as 
\begin{align}
 \uv[t] = \alpha \Bm^{-1} (\xv_0 - \hat{\xv}[t]). \label{eqn:control_vg}
\end{align}

\emph{Generator swing equations:} The swing equations establish a mathematical relationship between the angles of the mechanical motor and
the generated alternating current electricity \cite{Kundur1994}. The swing equations can be linearized and modeled 
as an LTI system described by Eqs. \eqref{eqn:process} and \eqref{eqn:Obs} under the assumption of direct current (DC) power flow\cite{FabioPow2011}. For a power network consisting of $n$ generators, the state vector consists of $2n$ entries. The first $n$ entries are the generator's rotor phase angles and
the last $n$ entries are the generator's rotor frequency. The control inputs correspond to 
changes in mechanical input power to the
generators, and is responsible for maintaining the generator's rotor angle and frequency within a safe operational range.
The entries of the matrix $\Am$ depend on the power system's topology (including the transmission lines' susceptances) 
as well as the generators' mechanical parameters (such as inertia and damping constants). 
The structure of the matrix $\Bm$ depends on the type of feedback control used to restrict the rotor 
angle frequency to within the safety range \cite{Kundur1994}. 
The measurement vector $\yv[t]$ under the DC power flow model includes nodal real power
injections at all the buses, all the branch power flows, and the rotor angles. The observation matrix $\Cm$ can be constructed based on the power system topology \cite{Liu2009}.

\subsection{Threat Model, Attack Detection \& Mitigation}
Modern-day critical infrastructure systems extensively use ICT for their operation.
For instance, in a power grid, the remote terminal units (RTUs) and many other field
devices are connected by the internet protocol (IP). The sensor and
control data is transmitted over the Internet using virtual private networks (VPNs) for 
logical isolation \cite{Hahn2013}. 
However it has been demonstrated in the past that software-based protection schemes such as VPNs can be breached by attackers (e.g., see \cite{Heartbleed}).
Additionally, in a power grid, the sensors (such as the voltage and current measurement units) are spread over a large geographical area, making their measurements vulnerable to physical attacks \cite{kune2013ghost, SmartMeterSecurity2009}. 
Such vulnerabilities can be exploited to launch attacks and
disrupt the normal power grid operations. 

In this paper, we follow Kerckhoffs's principle and consider an 
attacker who has accurate knowledge of the targeted CPCS and read access to the system state.
Such knowledge can be obtained in practice by malicious insiders, long-term
data exfiltration \cite{dragonfly2014}, or social engineering against employees,
contractors, or vendors of a critical infrastructure operator \cite{karnouskos2011}.
Specifically, we assume that the attacker knows the matrices $\Am, \Bm$ and $\Cm,$ as well as the operational details of the KF and the system's method of anomaly detection (including the detection threshold). In addition, the attacker also has read and write 
accesses to the system sensors.

We consider FDI attacks on the system sensors. Under this attack model, the compromised
observations, denoted by $\yv_a[t]$, are given by
\begin{align}
\yv_a[t] &= \yv[t] + \av[t] \label{eqn:Obs_Attack},
\end{align}
where $\av[t] \in \RR^{m}$ is the attacker's injection.
To model the attacker's energy constraint, we assume that the norm of the injection, $\|\mathbf{a}[t]\|$, is upper-bounded by a constant $a_{\max}$, i.e., $||\av[t]|| \leq a_{\max}.$
Denote by $\mathcal{A}$ the set of all feasible attack vectors that satisfy the above energy constraint. 

We assume that the controller uses the $\chi^2$ detector \cite{MEHRAChiSquare1971} to detect the attack, which
has been widely adopted in security analysis of LTI systems \cite{Kwon2013}, \cite{Mo2015}. We note that our analysis framework can also be extended to address other attack detectors.
The $\chi^2$ detector computes a
quantity
$g[t] = \rv[t]^T \Pm^{-1}_r \rv[t],$
where $\rv[t]$ is the residual given by
\begin{align}
\rv[t] &= \yv_a[t+1] - \Cm (\Am \hat{\xv}[t] + \Bm \uv[t]) \label{eqn:res_defn},
\end{align}
and $\Pm_r = \Cm \Pm_{\infty} \Cm+\Rm$ is a constant matrix that denotes the covariance of the residual in the steady state. Denoted by $i[t] \in \{0, 1\}$ the detection result of the detector. The detector declares an attack if $g[t]$ is greater than a predefined threshold $\eta.$ 
Specifically,
\begin{align}
i[t] = 
\begin{cases}
	0, & \text{if} \ 0 \leq g[t] \leq \eta; \\
	1, & \text{else}.
\end{cases} \label{eqn:indicator}
\end{align}

Based on the detection result, the controller applies a reactive mitigation action. 
If the $\chi^2$ detector's alarm is triggered, the controller forwards a modified version of the observation $\yv_a[t] - \deltav[t]$  
to the KF, where $\deltav[t] \in \RR^m$ is an attack mitigation signal;
otherwise, the controller directly forwards $\yv_a[t]$ to the KF (ref. Fig.~\ref{fig:sys_model}). 
Thus, the controller's operation can be expressed as
\begin{align}
\yv_f[t] = \yv_a[t] - i[t] \deltav[t]. \label{eqn:mitigation}
\end{align}
With the controller's mitigation action, the KF estimate is computed as
\begin{align}
\hat{\xv} [t+1] \!=\! \Am \hat{\xv}[t]\! +\! \Bm \uv[t] \! +\! \Km (\yv_f [t+1]\! -\! \Cm (\Am \hat{\xv}[t]\! +\! \Bm \uv[t])). \label{eqn:KF_est}
\end{align}

The mitigation signal $\deltav[t]$ can be generated using existing mitigation approaches (e.g., \cite{Cardenas2011}, \cite{Sridhar2014}).
The main focus of this paper is not the design of the mitigation strategy, but to understand 
the impact of the detection-mitigation loop on the optimal attack strategy.
Thus, in this paper, we do not focus on a specific mitigation approach. Instead, we design a generic framework 
that admits any mitigation signal.
In Section~\ref{sec:Sim_Res}, our simulations are based on a perfect mitigation strategy in which the
controller can precisely remove the attack signal, as well as a practical mitigation strategy in which the mitigation signal is a noisy version of the attack signal.

Combining \eqref{eqn:process}, \eqref{eqn:mitigation} and \eqref{eqn:KF_est}, 
we obtain the dynamics of the KF estimation error with attack mitigation as
\begin{align}
& \ev  [t+1]  =  \Am_K \ev[t] + \Wm_K\wv [t]  \nonumber \\  & \ -  \Km (\av[t+1]-i[t+1] \deltav[t+1])   - \Km \vv[t+1], \ t \geq 0,  \label{eqn:error_evol}
\end{align}
where $\Am_K = \Am- \Km \Cm \Am$ and $\Wm_K = (\Id- \Km \Cm).$
Since the KF
is assumed to be in the steady state at time $0,$ we have $\mathbb{E}[\ev[0]] = \bf{0}$ and $\mathbb{E} [\ev[0] \ev[0]^T] = \Pm_e = (\Id- \Km \Cm) \Pm_\infty.$

\section{Problem Formulation}
\label{sec:Threat_Model}
Under the Kerckhoffs's assumption about the attacker's
knowledge, we analyze attack strategies that can mislead the controller into making erroneous control
decisions. This is accomplished indirectly by 
increasing the estimation errors. 
For a given attack detection threshold $\eta$ and mitigation strategy $\{ \deltav[t] \}^T_{t = 1}$ over a horizon of $T$ time slots, the optimal attack sequence that maximizes the cumulative sum of KF's expected norm of the estimation error over the horizon is given by the following optimization problem:
\beqa
&\dsp  \max_{ \av[1],\dots,\av[T]} &  \sum^T_{t = 1}\mathbb{E} [\| \ev[t] \|^2 ] \label{eqn:attacker_problem} \\ 
& s.t. &  \text{KF error dynamics \eqref{eqn:error_evol}}, \nonumber \\
& & \|\mathbf{a}[t] \| \leq a_{\max}, \forall t \nonumber.
\eeqa 
Maximizing the KF estimation error implies that the controller no longer has
an accurate estimate of the system state. 
In systems that use KF for state estimation (such as positioning systems, power systems, etc.), control input computed based on inaccurate/wrong system state estimates can adversely affect their performance and even result in catastrophic safety incidents. 
Moreover, the cumulative sum in the objective function implies that the attack has a sustained adverse impact on the system over the entire attack time horizon. 
We note that similar cumulative metrics have also been widely adopted in control system design to assess the performance of controllers \cite{abdelzaher2008introduction}. Thus, with an objective of maximizing the cumulative metric, the optimal attack sequence will bring the largest performance degradation to the control systems that are designed in terms of cumulative metrics.

\begin{figure}[!t]
\centering
\includegraphics[width=0.48\textwidth]{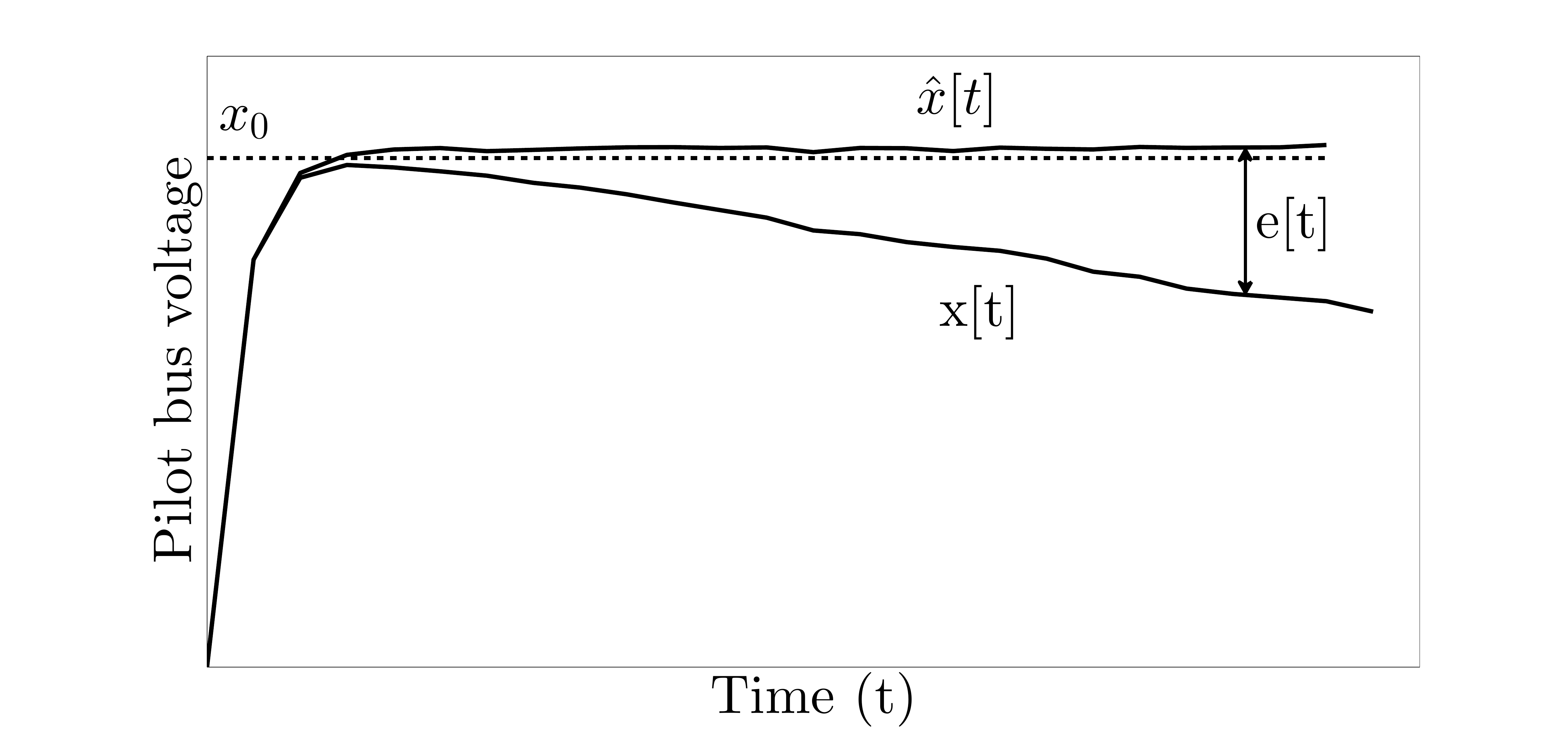}
\caption{Attack impact for the voltage control problem.}
\label{fig:obj_fn}
\end{figure}

\subsubsection*{Relevance to Power System}
We illustrate the relevance of the optimization problem
stated in \eqref{eqn:attacker_problem} to power grid's voltage control. Recall that the voltage controller's objective is to adjust the pilot bus voltage to its setpoint $\xv_0$ by applying control.
Fig.~\ref{fig:obj_fn} shows the impact of an attack that is able to bypass the  $\chi^2$ detector (and consequently the controller's mitigation steps) on the pilot bus voltage. 
In this figure, the dotted line indicates the voltage setpoint, and the solid lines show the evolution of the system state $\xv[t]$ and estimate $\hat{\xv}[t]$. The gap between the two curves measures the KF estimation error $\ev[t].$ As evident from the figure, if the attacker manages to increase the KF's estimation error using a carefully constructed attack sequence, then he can cause a significant deviation of the system state from the desired setpoint. Interestingly, the estimate $\hat{\xv}[t]$ is close to the setpoint ${\xv}_0$ that misleads the controller into believing that the desired setpoint has already been achieved, while the actual pilot bus voltage continues to deviate.

Intuitively, to cause a significant impact, the attack magnitude must be large. But at the same time, it is important that the attack bypasses the controller's detection -- otherwise the attack will be mitigated. Thus the solution of the optimization problem \eqref{eqn:attacker_problem} must strike a balance between the attack magnitude and stealthiness. 
In the following section, we solve the optimization problem \eqref{eqn:attacker_problem} 
using an MDP-based approach.

\section{MDP Solution}
\label{sec:Soln_Methods}
In this section, we cast the optimization problem \eqref{eqn:attacker_problem} to an MDP problem \cite{Puterman:1994} and solve it using the value iteration method. 
Before doing so, we first state the main challenge involved in solving \eqref{eqn:attacker_problem}.

\subsection{Challenge}
\begin{figure}[!t]
\centering
\includegraphics[width=0.45\textwidth]{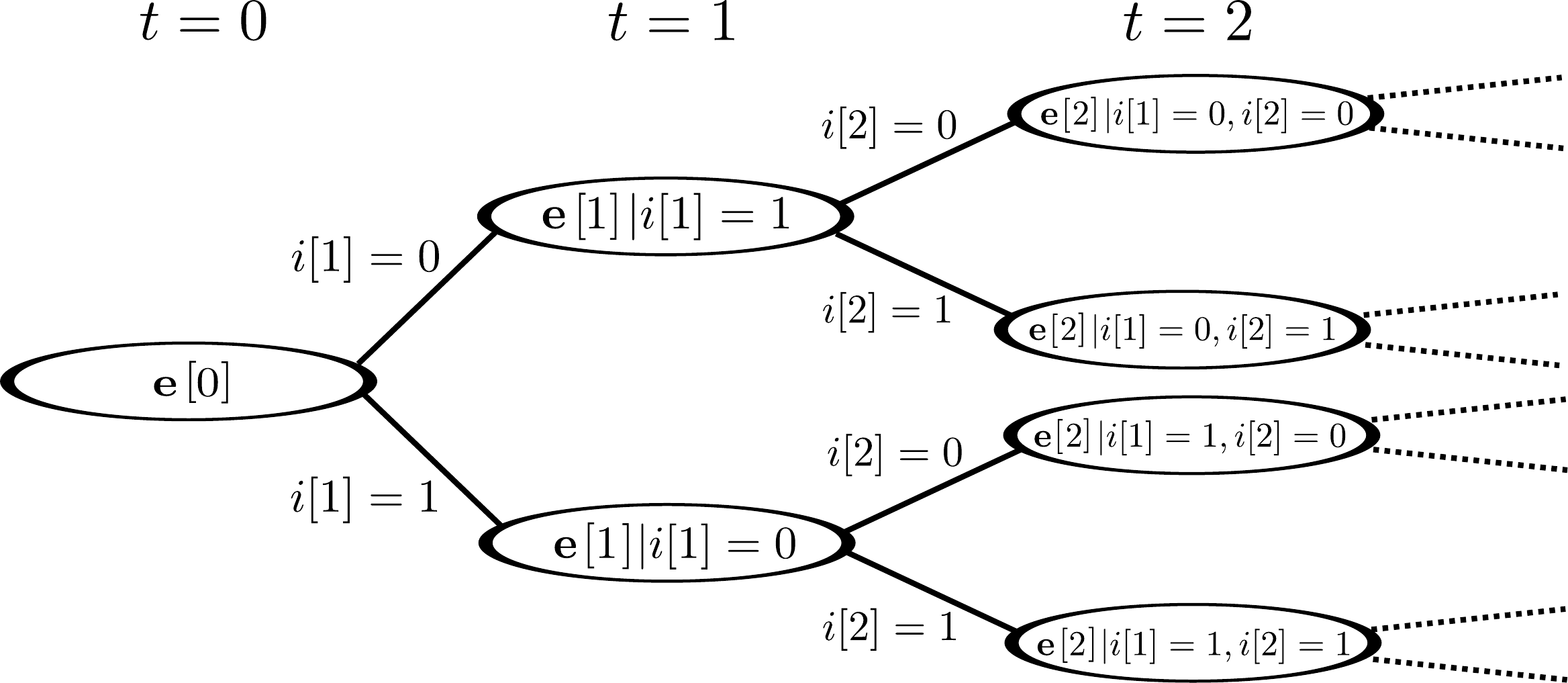}
\caption{Evolution of KF estimation error conditioned on the attack detection results.}
\label{fig:probabilistic_evolution}
\vspace*{-0.5 cm}
\end{figure}

The main challenge in solving \eqref{eqn:attacker_problem} lies in the fact that 
the KF error dynamics, and consequently the attack detection results are coupled
across different time slots.  
To illustrate this point, we present a pictorial depiction of the KF
error dynamics \eqref{eqn:error_evol} in Fig.~\ref{fig:probabilistic_evolution}. 
As evident from this figure, the error dynamics of $\ev[t]$ depend on 
the sequential decisions of the $\chi^2$ detector $\iv_{[1:t]} =  \{ i[t] \}^T_{t = 1}$
due to reactive attack mitigation, which is triggered based on the the attack detection.
Thus, to compute the expected error at any time $t$, the attacker must consider
all possible combinations of the past attack detection results $\iv_{[1:t]} =  \{ i[t] \}^T_{t = 1}$. 
The complexity of such an approach grows exponentially in terms of the optimization time horizon $T$ (since at any time $t,$ there can be $2^t$ different combinations of the past attack detection results, see Fig.~\ref{fig:probabilistic_evolution}). In the following subsections, we present an efficient solution methodology to solve the attacker's problem \eqref{eqn:attacker_problem} by modeling it as an MDP, and propose a value iteration based method to compute the optimal attack sequence.

\subsection{Markov Decision Process Model}
\label{sec:MDP_Model}
In this subsection, we develop the MDP modelling of the optimization problem \eqref{eqn:attacker_problem}.
Our key observation is that the dynamics of the KF estimation error in \eqref{eqn:error_evol} is Markovian. Hence, the knowledge of $\ev[t]$ at time $t$ will capture all the past events, and exhaustive search across all the possible past attack detection results is not necessary.

A state in the MDP corresponds to the
KF filter estimation error $\ev[t]$ and the actions correspond 
to the attacker's injection $\av[t].$ 
Our approach is to map the KF error dynamics \eqref{eqn:error_evol} to the
state transition probabilities of the MDP, and the objective function of \eqref{eqn:attacker_problem} to the MDP's long-term expected reward. The solution to the MDP is a policy which maps
each MDP state to an attacker's action. In particular, the optimal policy maximizes
the long-term expected reward of the MDP, and hence solves the optimization 
problem \eqref{eqn:attacker_problem}.
The mathematical details of the MDP is presented next. The structure of
the MDP's solution is illustrated with the help of a numerical example in Section~\ref{sec:Mag_Stl}.

\subsubsection*{MDP Modeling Details}
Formally, the MDP is defined by a tuple $(\mathcal{E},\mathcal{A},\mathcal{T},R)$, where
$\mathcal{E} \subseteq \RR^n$ is the state space of the problem corresponding 
to the set of all possible $\ev[t]$.  $\mathcal{A}$ is the action space of the attacker.  $\mathcal{T}(\ev,\av,\ev^{\prime})$ is the probability of transition from state
$\ev$ to $\ev^{\prime}$ (where $\ev,\ev^{\prime} \in \mathcal{E}$) under an action $\av \in \mathcal{A}$ of the attacker. Mathematically,
$\mathcal{T}( \ev,\av,\ev^{\prime})  \defines \mathbb{P}  (\ev[t+1] = \ev^{\prime} \big{|} \ev[t] = \ev,\av[t+1] = \av) .$
$R(\ev^{\prime},\av,\ev)$ is the immediate expected reward for the attacker when it takes an action 
$\av \in \mathcal{A}$ in state $\ev \in \mathcal{E}.$

\emph{MDP state transition probabilities:} We now compute the state transition probability corresponding to the error dynamics \eqref{eqn:error_evol}. 

We adopt the following approach. First, we compute the quantity $\mathbb{P} ( \ev_{\text{lb}} \leq \ev[t+1]  \leq \ev_{\text{ub}} \big{|} \ev[t] = \ev,\av[t+1] = \av).$ 
Then we use the fact that for a random variable $X,$ 
\begin{align*}
& \mathbb{P}  (X = x) \approx \frac{F ( -\infty ,x+\epsilon) - F ( -\infty ,x-\epsilon) }{2 \epsilon},
\end{align*}
where $F ( x_1 ,x_2) = \mathbb{P} (  x_1 \leq X  \leq x_2)$ and $\epsilon > 0$ is a small positive quantity.

The result is stated in the following lemma:
\begin{lemma}
\label{lem:trans_prob}
For a given $\ev[t] = \ev$ and $\av[t+1] = \av$ the attack detection probability at
any time $t$ can be computed as $\mathbb{P} ( \Ym \geq \eta),$ where $\Ym = \rv_c[t+1]^T \Pm^{-1}_r \rv_c[t+1]$ is a generalized chi-square distributed random variable. 
Further, the quantity $\mathbb{P} ( \ev_{\text{lb}} \leq \ev[t+1]  \leq \ev_{\text{ub}} \big{|} \ev[t] = \ev,\av[t+1] = \av)$  can be computed as the sum of the following terms:
\begin{align}
& \mathbb{P} \LB \begin{bmatrix}
 0  \\
 \ev_{\text{lb}} - \yv_2
\end{bmatrix}  \leq \Xm   \leq \begin{bmatrix}
 \eta  \\
 \ev_{\text{ub}} - \yv_2
\end{bmatrix} \RB \nonumber \\ 
& + \mathbb{P} \LB \begin{bmatrix}
 \eta  \\
 \ev_{\text{lb}} - \yv_2 - \Km \deltav
\end{bmatrix}  \leq \Xm   \leq \begin{bmatrix}
 \infty  \\
 \ev_{\text{ub}} - \yv_2 - \Km \deltav
\end{bmatrix} \RB.
\label{eqn:TP_vector}
\end{align}
In \eqref{eqn:TP_vector}, $\Xm \in \RR^{n+1}$ is a concatenated variable given by \\
$\Xm  = \LSB \Ym \ \ (\Wm_K \wv[t]- \Km \vv[t+1] )^T \RSB^T,$ $\yv_2 = \Am_K\ev - \Km\av,$
and $\deltav$ is the mitigation signal.
\end{lemma}
Lemma~\ref{lem:trans_prob} is proved in Appendix~A.
For a generic system of dimensions $n, m \geq  2,$ it is hard to obtain analytical expressions for the probability terms involved in Lemma~\ref{lem:trans_prob} (since they involve a generalized chi-square distribution, as well as the correlations between the random variables $\Ym$ and $\Wm_K \wv[t]- \Km \vv[t+1],$ which is hard to quantify analytically).
However, for the scalar case i.e. $n = m = 1$, the attack detection and transition probabilities
can be computed using the Gaussian distribution, as stated in the following corollary:
\begin{corollary}
\label{cor:trans_prob}
For $n = m = 1,$ the attack detection probability at
any time $t$ can be computed as
\begin{align}
& \mathbb{P} \big{(}\Ym \in  (-\infty, -\sqrt{\eta \Pm_r} - \Cm\Am\ev - \av] \nonumber \\ & \qquad \qquad \qquad \qquad \cup [\sqrt{\eta \Pm_r} - \Cm\Am\ev - \av, \infty) \big{)}, \label{eqn:Det_Prob}
\end{align}
where $\Ym \sim \mathcal{N} (0,\Cm \Qm \Wm^T_K + \Rm).$
Further, the quantity $\mathbb{P} ( \ev_{\text{lb}} \leq \ev[t+1]  \leq \ev_{\text{ub}} \big{|} \ev[t] = \ev,\av[t+1] = \av)$ is equal to the sum of the following terms: 
\begin{align}
& \mathbb{P} \LB \begin{bmatrix}
-\sqrt{\eta \Pm_r} - \yv_1 \\
\ev_{\text{lb}} - \yv_2
\end{bmatrix} \ \leq \Xm \leq 
\begin{bmatrix}
\sqrt{\eta \Pm_r} - \yv_1 \\
\ev_{\text{ub}} - \yv_2
\end{bmatrix} \RB \nonumber \\
+
& \mathbb{P} \LB \begin{bmatrix}
-\infty \\
\ev_{\text{lb}} - \yv_2 - \Km \deltav
\end{bmatrix} \ \leq \Xm \leq 
\begin{bmatrix}
-\sqrt{\eta \Pm_r} - \yv_1 \\
\ev_{\text{ub}} - \yv_2- \Km \deltav
\end{bmatrix} \RB \nonumber \\ 
+
& \mathbb{P} \LB \begin{bmatrix}
\sqrt{\eta \Pm_r} - \yv_1 \\
\ev_{\text{lb}} - \yv_2- \Km \deltav
\end{bmatrix} \ \leq \Xm \leq 
\begin{bmatrix}
\infty \\
\ev_{\text{ub}} - \yv_2 - \Km \deltav
\end{bmatrix} \RB \label{eqn:norm_cdf}
\end{align}
where $\yv_1 = \Cm\Am\ev + \av$ and $\yv_2 = \Am_K\ev - \Km\av$ and $\Xm \in \RR^{2}$ is a zero-mean Gaussian distributed random vector whose covariance matrix is given by
\begin{align}
\text{Cov} & (\Xm)   = \begin{bmatrix}
\Cm \Qm \Wm^T_K + \Rm  & \Cm \Qm \Wm^T_K  - \Rm \Km^T \\
\Wm^T_K \Qm \Cm  - \Km \Rm^T & \Wm_K \Qm \Wm^T_K+ \Km \Rm  \Km^T
\end{bmatrix}.
\end{align}
\end{corollary}
The probabilities in \eqref{eqn:Det_Prob} and \eqref{eqn:norm_cdf} can be computed using the cumulative distribution function (c.d.f.) of Gaussian distribution. Corollary~\ref{cor:trans_prob} is also proved in Appendix~A.

\emph{MDP reward:} We now map the objective function of \eqref{eqn:attacker_problem} to the 
MDP reward function. Accordingly, the immediate expected reward of the MDP is given by
\begin{align}
R(\ev^{\prime},\av,\ev) = \int_{\ev^{\prime} \in \mathcal{E}}\mathcal{T}( \ev,\av,\ev^{\prime}) || \ev^{\prime} ||^2. \label{eqn:imm_rew}
\end{align}

\emph{MDP policy and state value function: }The solution to the MDP corresponds to a policy $\pi,$ which is a mapping from a state to an action. 
The state value function of the MDP for a given policy $\pi$ is defined as
\begin{align}
V^{\pi} (\ev) = \mathbb{E}_{\pi} \LSB \sum^T_{t = 1} ||\ev[t]||^2 \big{|} \ev[0] = \ev \RSB.
\end{align}
\emph{Optimal policy: }The optimal policy
$\pi^*$ maximizes the total expected reward, $
\pi^* = \argmax_{\pi} V^{\pi} (\ev) , \forall \ev \in \mathcal{E},$ and
the optimal value function is defined as $V^*(\ev) = V^{\pi^*} (\ev).$

In the next subsection, we present an algorithm to compute the optimal policy of the MDP described above.

\subsection{Solving the MDP}
\label{sec:MDP_Discrete}
MDPs can be solved efficiently by value/policy iteration methods \cite{Puterman:1994}. However, 
in this work we are dealing with real-world quantities (for e.g. voltages in a
power grid) which are continuous variables. Hence, the MDP described in Section~\ref{sec:MDP_Model} has continuous state and action spaces\footnote{We note that the MDP problem has continuous state and action spaces, but is not a continuous-time MDP (since we only consider discrete-time LTI systems).}, which makes it impractical to apply a value iteration method directly. In order to address this issue, in what follows, we
define a \emph{discretized MDP} obtained by discretizing the state space of the original continuous MDP.
The optimal policy of the discretized MDP can be used as a near-optimal solution to the continuous MDP. 
Existing studies (e.g., \cite{ChowTsitsiklis1991}) adopt similar discretization approaches.
In the following, we provide only a sketch of the discretization procedure. More details of the discretized procedure can be found in Appendix B. This is followed by a value iteration algorithm to compute its optimal policy.

The MDP discretization procedure is based on the following three steps: 
\begin{itemize}
\item[1.] Construct a discretized MDP that mimics the continuous MDP closely.
\item[2.] Solve the discretized MDP using value iteration, which gives an optimal
policy for the discretized MDP.
\item[3.] Map the discterized MDP's optimal policy to a near-optimal policy for the continuous MDP. 
\end{itemize}
Let $\Xi$ denote the discretized version of the original state space
$\mathcal{E},$ where $\Xi = \{ \xi_1,\dots,\xi_N \},$ 
where $N$ is the number descritization levels, and $\overline{\mathcal{T}} (\xi_i,\av,\xi_j)$, $\overline{R}(\xi_i,\av,\xi_j)$ and 
$\overline{V} (\xi_i)$ denote the state transition probabilities, the reward
and value function of the discretized MDP. The mathematical details of their computation 
are provided in Appendix~B. 
The discretized MDP can be solved using the value iteration method whose steps
are given by the following algorithm:
\vspace{0.05in} \hrule
\vspace{0.01in}\hrule
\begin{algorithm} [{\bf Value Iteration}]
\label{alg:HeurisitcSingleStorage}
\begin{algorithmic}[1] 
\item[]
\State Set $\overline{V}^*_{0} (\xi_i) = 0$ for all $\xi_i \in \Xi.$
\For{ t = 0 to T-1}
\For{ all discretized states $\xi_i \in \Xi$}
\begin{align*}
& \overline{V}^*_{t+1} (\xi_i)  \\ & \ \ \leftarrow \max_{\av} \sum_{\xi_j \in \Xi} \overline{\mathcal{T}} (\xi_i,\av,\xi_j) \LSB \overline{R}(\xi_i,\av,\xi_j) + \overline{V}^*_t (\xi_j) \RSB, \\
& \overline{\pi}^*_{t+1}(\xi_i)  \\ & \ \ \leftarrow \argmax_{\av} \sum_{\xi_j \in \Xi} \overline{\mathcal{T}}(\xi_i,\av,\xi_j) \LSB \overline{R}(\xi_i,\av,\xi_j) + \overline{V}^*_t (\xi_j) \RSB.
\end{align*}
 \EndFor
 \EndFor
\end{algorithmic}
\end{algorithm}
 \hrule
\vspace{0.01in}\hrule\vspace{0.05in}
\vspace*{0.1 in}
\noindent Algorithm~1 gives the optimal policy of the discretized MDP \cite{Puterman:1994}.

Note that the optimal policy of the discretized MDP computed in Algorithm~1 cannot be
directly applied to the continuous MDP, since we do not know the optimal policy 
for a state $\ev \in \mathcal{E}$ that is not in the discretized state space $\Xi.$
To address this issue, we use the nearest neighbour approximation, i.e.,
for a state $\ev \notin \Xi,$ we choose an action based on
the policy of its nearest neighbour,
$\pi(\ev) = \bar{\pi}^*(\xi_i), \ \text{where} \ \xi_i = \argmin_{1 \leq i \leq N} ||\ev-\xi_i||$.
We lastly make some remarks on the MDP formulation in this section.

$\bullet$ Although in this section we cast the optimization problem \eqref{eqn:attacker_problem} as a finite time horizon MDP problem, our framework can be extended to the infinite time horizon MDP problem readily by introducing a discount factor $0 \leq \gamma < 1$ in the reward function.
The discount factor ensures that the cumulative sum of rewards is finite as well as the convergence of the value iteration algorithm.

$\bullet$ The optimal cost of the discretized MDP 
is guaranteed to lie within a bounded distance from the optimal cost of the original MDP 
\cite{ChowTsitsiklis1991}. 
As the discretization is finer, the discretized MDP approaches to the original
MDP more closely.

\subsection{Attack Magnitude and Stealthiness}
\label{sec:Mag_Stl}
\begin{figure}[!t]
\centering
\begin{subfigure}{0.23\textwidth}
\includegraphics[width=1\textwidth]{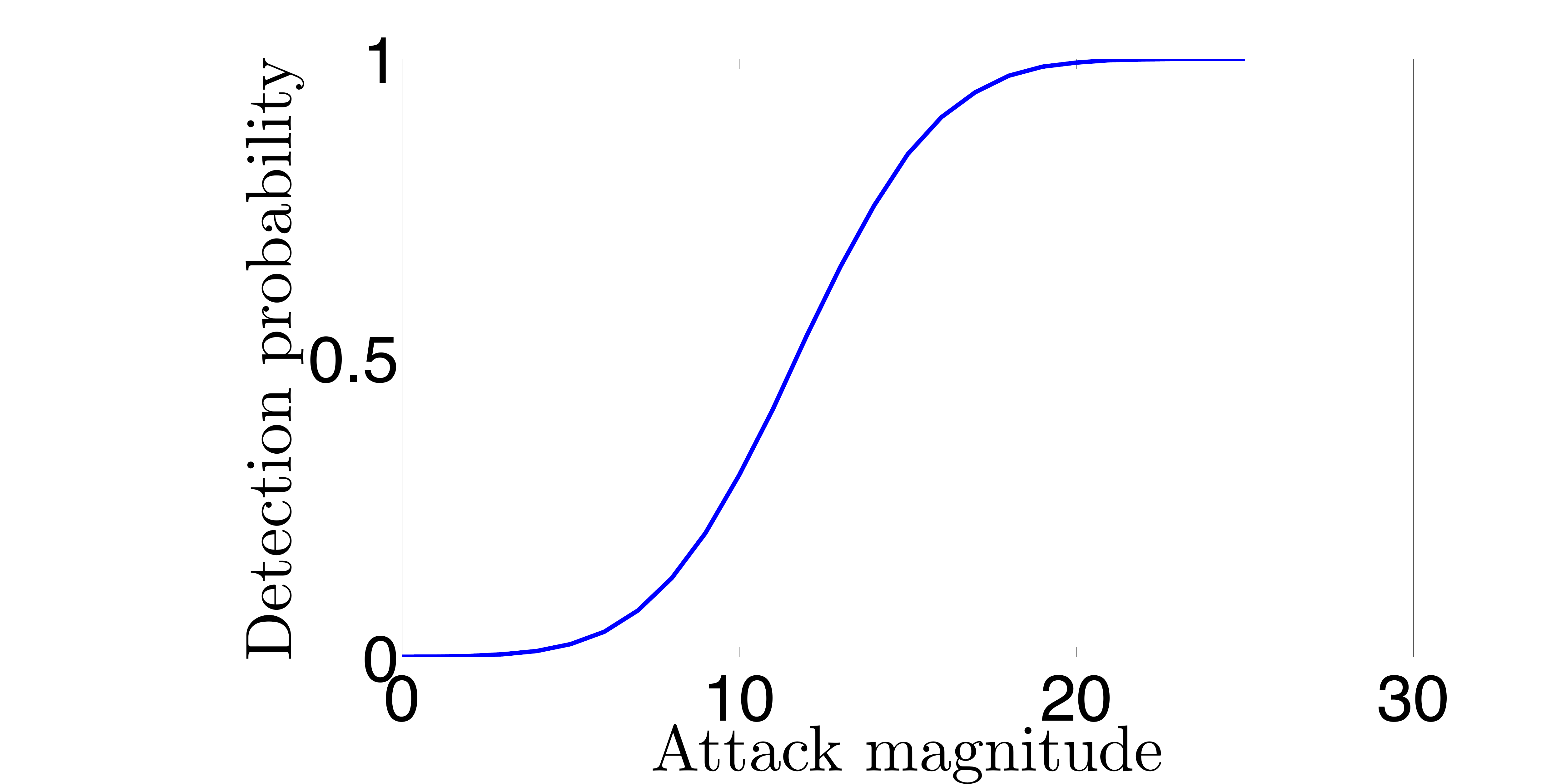}
\end{subfigure}
~
\begin{subfigure}{0.23\textwidth}
\includegraphics[width=1\textwidth]{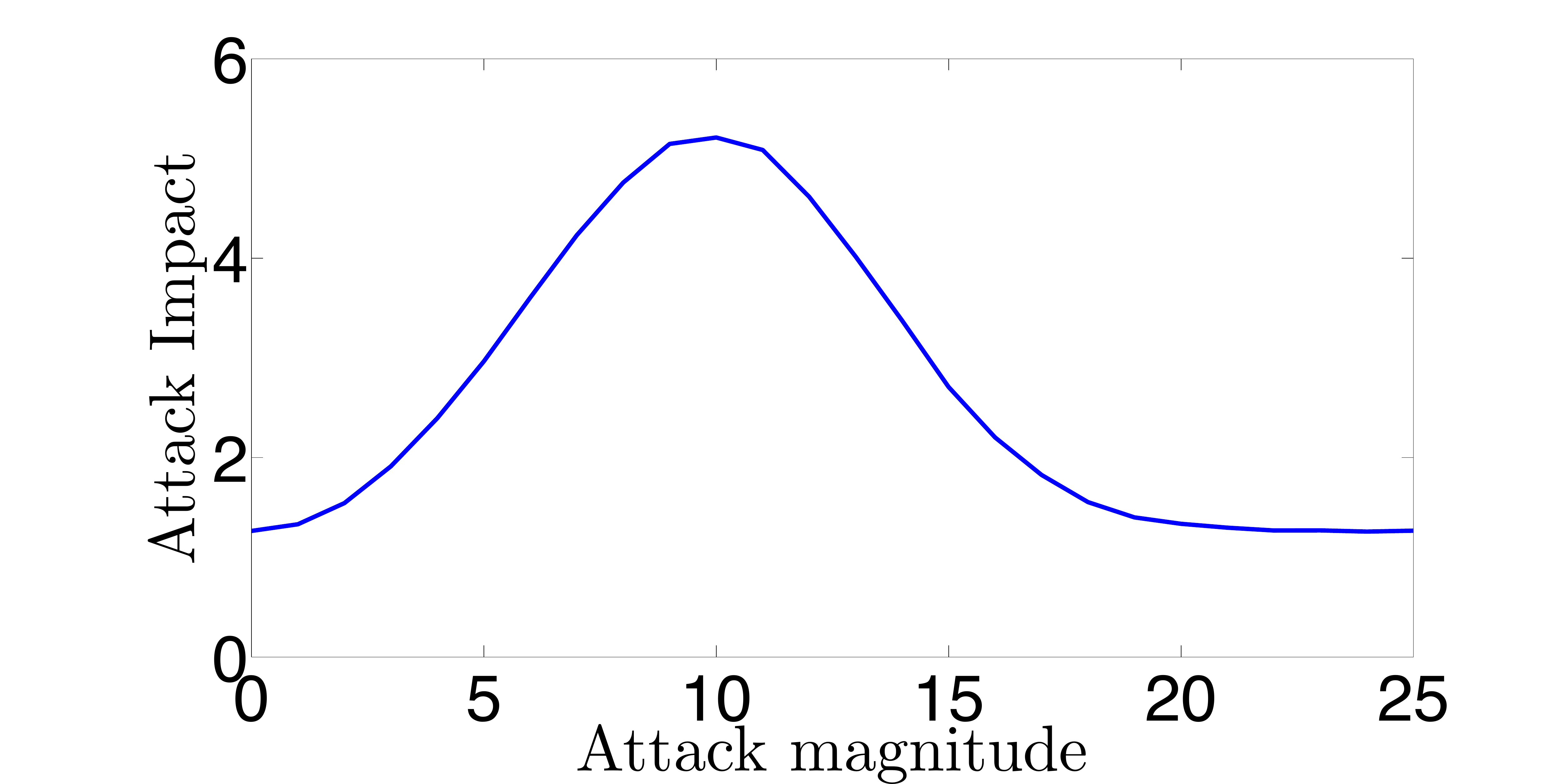}
\end{subfigure}
\caption{Attack detection probability and the expected attack impact (immediate expected reward of MDP) for different attack magnitudes.}
\label{fig:Attack_Mag_Stl}
\end{figure}

We now illustrate the structure of the MDP solution using a numerical example.
In Fig.~\ref{fig:Attack_Mag_Stl}, we plot the attack detection probability (computed as in \eqref{eqn:Det_Prob}) and the attack impact computed
in terms of the MDP's immediate expected reward (using the result of \eqref{eqn:norm_cdf} and \eqref{eqn:imm_rew}) for different values of attack magnitude $\av$. The system parameters are $n = m = 1,$ $\Am$ = $1$, $\Cm$ = $1$, $\Qm$ = $1$, $\Rm$ = $10,$  $\eta = 10$ and $\deltav = \av$. It can be observed that while the probability of detection is low for an attack of small magnitude, it also has little impact. On the other hand, the probability of detection is high for an attack of large magnitude, and consequently the expected attack impact is also low. The optimal attack lies in between these two quantities. In this example, the optimal attack that maximizes the expected immediate reward has a magnitude of $10,$ and 
a detection probability of $0.3.$ Thus, the MDP solution strikes a balance
between attack magnitude and stealthiness, resulting in maximum impact\footnote{Strictly speaking, MDP solution maximizes the long term expected reward. For the ease of illustration, in this example we only considered the immediate expected reward.}.

\section{Cost of False Positives and Misdetections}
\label{sec:MD_FP}
In this section, we use the framework developed thus far to quantify 
the cost of FPs and MDs in a simulation-based approach.
We use the cumulative state estimation error (objective function of \eqref{eqn:attacker_problem}) 
as the cost metric.

To quantify these costs, we consider an LTI system with an  \emph{oracle} attack detector 
as the reference system. An oracle detector is one that has a perfect detection capability, and hence no FPs or MDs. The cost of FPs is the additional cost incurred due to wrongly triggered mitigations in the original LTI system, compared with the reference system. The cost of MDs is the additional cost incurred due to unmitigated attacks in the original LTI system, compared with the reference system. In particular, we consider optimal attacks as derived in Section~\ref{sec:MDP_Model} to characterize the worst-case performance degradation due to MDs.
We compute these costs as follows:

{\bf Cost of FPs:} To quantify the cost of FPs, we compute the state estimation error (objective function of \eqref{eqn:attacker_problem}) in the following two systems: (i) the LTI system of \eqref{eqn:process} and \eqref{eqn:Obs} with the $\chi^2$ detector and mitigation modules and no attacks, i.e. $\av[t] = 0, \ \forall t$ (ii) the reference LTI system with $\av[t] = 0, \ \forall t$.

Under setting (i), all the alarms of the $\chi^2$ detector correspond to FPs, which will wrongly trigger a mitigation action. Since the mitigation signal is imperfect, it leads to an increase in the estimation error. Note that for the reference system, there are no FPs, and hence no wrongly triggered mitigations. The difference between the state estimation errors of the two systems quantifies the performance degradation due to FP.

{\bf Cost of MDs:} 
To quantify the cost of MDs, we compute the state estimation errors in the following two systems: 
(i) the LTI system of \eqref{eqn:process} and \eqref{eqn:Obs} with the $\chi^2$ detector and mitigation modules and optimal attacks (computed as in Section~\ref{sec:MDP_Model}) (ii) the reference LTI system with optimal attacks. The difference between the state estimation errors of the two systems quantifies the performance degradation due to MD, which we define as the cost of MD. 

In Section~\ref{sec:quant_MD_FP}, we present simulation results to quantify the cost of FP and MD under different attack detection thresholds and mitigation strategies. We also provide
guidelines to tune the attack detection threshold based on this quantification.

\section{Simulation results}
\label{sec:Sim_Res}
In this section, we present simulation results to examine the system performance under 
different attack sequences. 
Throughout this section, we use different notations to denote the attacker's knowledge of the detection and mitigation parameters ($\eta_a$ and $ \{ \deltav_a[t] \}^T_{t=1}$, respectively), and the actual parameters used by the controller ($\eta_d$ and $\{ \deltav_d[t]\}^T_{t=1}$, respectively). While solving the attacker's problem \eqref{eqn:attacker_problem}, we assume \emph{perfect attack mitigation} in which the attack can be removed precisely, i.e. $\deltav_a[t] = \av[t], \ \forall t$. From an attacker's perspective, this assumption gives
an underestimate of the performance degradation he can cause (since the value of 
the objective function of \eqref{eqn:attacker_problem} will increase if the controller uses a mitigation strategy different from perfect mitigation).
 
While evaluating the attack impact, we consider two mitigation strategies used by the controller. First, the perfect attack mitigation $\deltav_d[t] = \av[t], \ \forall t.$ 
However perfect mitigation requires the controller to estimate the injected attack vector accurately, which may not be practical. Thus we introduce a \emph{practical attack mitigation approach} under which the attack mitigation is imperfect, i.e., $\deltav_d[t] = \av[t]+\bv[t],$ where $\bv[t] \in \RR^n$ models the mismatch between the controller's mitigation action and the actual attack vector (possibly due to  inaccuracy in estimating the injected attack). In our simulations, we generate a random vector to model $\bv[t]$.

\subsection{Optimality of the Attack Sequence}

First, we verify the optimality of the attack sequence derived using the MDP-based methodology described in Section~\ref{sec:Soln_Methods}. We consider a general LTI model described by \eqref{eqn:process} and \eqref{eqn:Obs} with $n = 1,$ $\Am$ = $1$, $\Cm$ = $1$, $\Qm$ = $1$, $\Rm$ = $10.$

We compare the cost function of \eqref{eqn:attacker_problem} under three different attack sequences:  (i) optimal attack computed using the MDP-based approach of Section~ \ref{sec:Soln_Methods}, (ii) a constant attack sequence of magnitude $10$ units and (iii) a ramp attack of the form $a[t] = t$. The time horizon of attack $T$ is fixed to $10$ units. To implement the discretized MDP, we truncate the state space in the range $[-30,30]$ and discretize it in equal intervals of $0.25$ units. Thus, the state space of the discretized MDP consists of a total of $241$ states, i.e., $\{ -30,-29.75,\dots,0,\dots,29.75,30\}$.
All the optimization problems involved in the implementation of the value iteration algorithm are solved using the \emph{fmincon} function in MATLAB. 


For attack impact, we compute an empirical value of the objective function of \eqref{eqn:attacker_problem} by conducting $W$ simulation runs (where $W$ is a large number). Let $\ev_{\omega}[t]$ denote the state estimation error at a time instant $t \in \{ 1,2,\dots,T \}$ during the simulation run $\omega = \{ 1,2,\dots,W\},$ where $\ev_{\omega}[t]$ follows the dynamics given by
\begin{align*}
& \ev_{\omega}  [t+1]  =  \Am_K \ev_{\omega}[t] + \Wm_K\wv [t]  \nonumber \\  & \ -  \Km (\av^*[t+1]-i[t+1] \deltav_d[t+1])   - \Km \vv[t+1], \ t \geq 0, 
\end{align*}
where $\av^*[t]$ is the attack derived from the MDP policy, i.e., $ \av^*[t] = \pi^*(\ev_{\omega} [t])$. The empirical cost at time $t$ is then computed by
averaging over the $W$ simulations, i.e.,
\begin{align}
\text{Cost}[t] = \frac{1}{W} \sum^W_{\omega = 1} \sum^t_{\tau = 1} \| \ev_{\omega}[t] \|^2. \label{eqn:emp_cost}
\end{align}
In our simulations, we set $W = 10000.$ To evaluate the empirical cost under other attack strategies,
we use a similar approach and replace the optimal attack with the corresponding attacks (i.e., constant and ramp attacks). 

\begin{figure}[!t]
\centering
\includegraphics[width=0.48\textwidth,trim={0 8cm 0 0}]{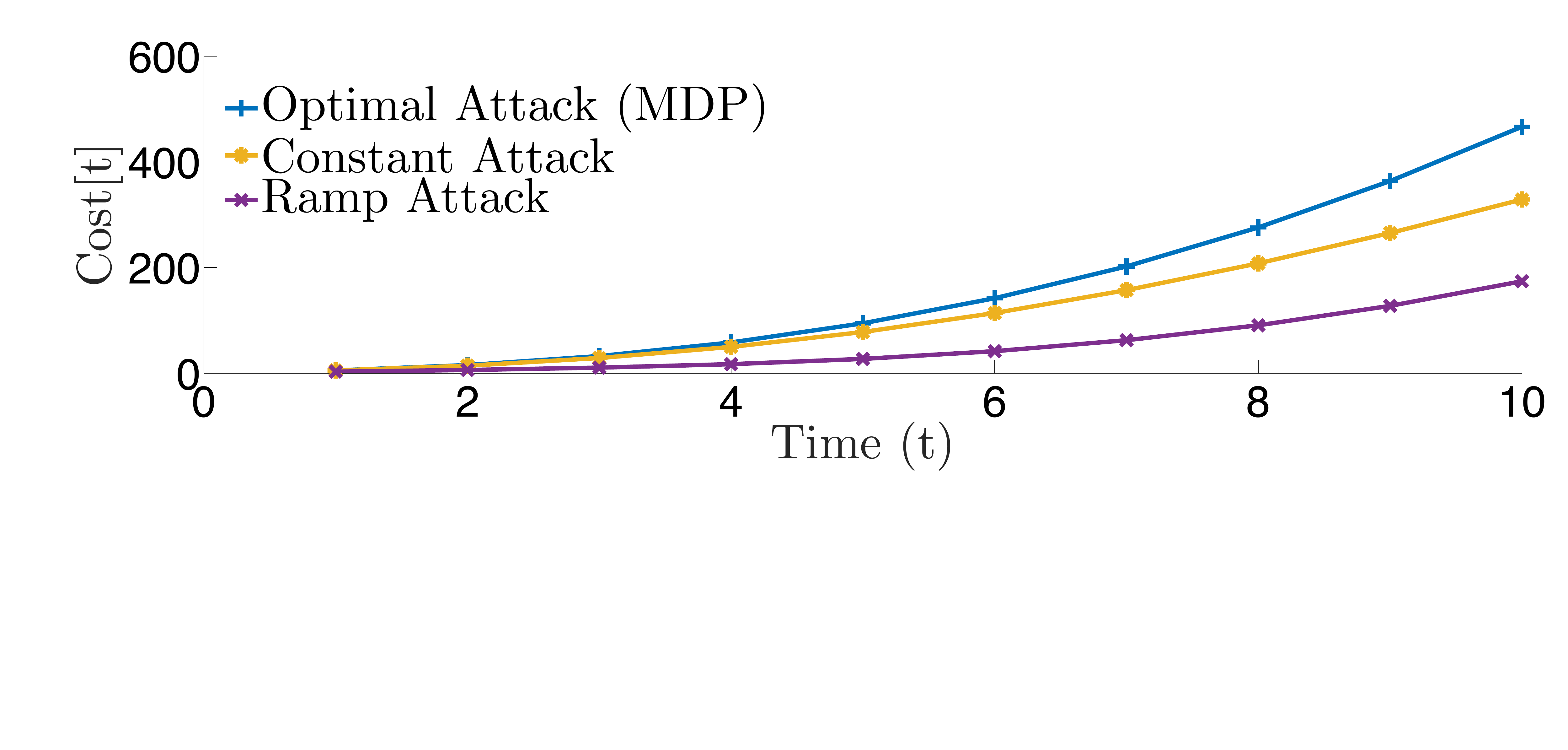}
\caption{Comparison of cost with perfect attack mitigation and different attack strategies.}
\label{fig:general_model_cost}
\end{figure}

\begin{figure}[!t]
\centering
\includegraphics[width=0.48\textwidth,trim={0 10cm 0 0}]{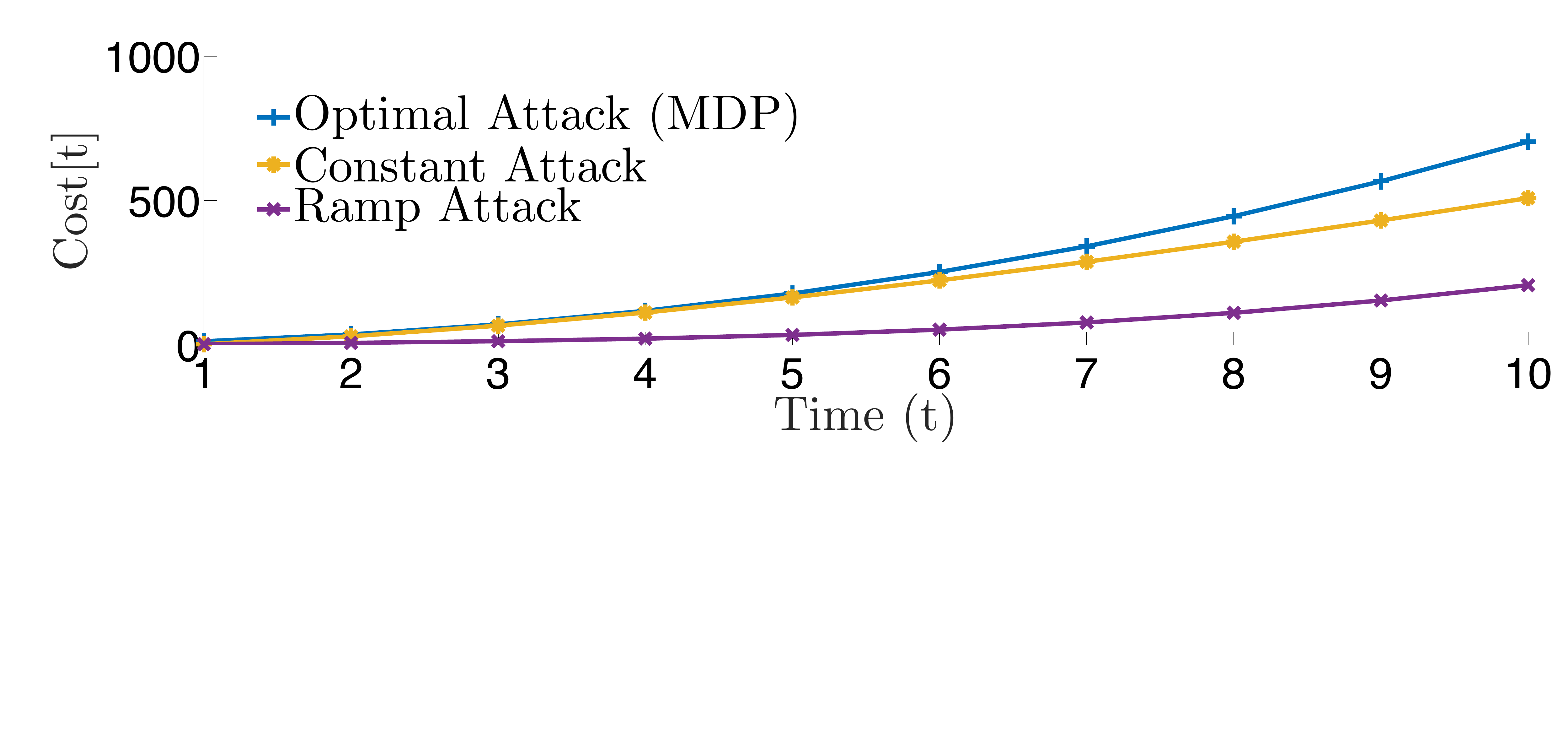}
\caption{Comparison of cost with practical attack mitigation and different attack strategies.}
\label{fig:cost_imperfect}
\end{figure}

Fig.~\ref{fig:general_model_cost} provides a comparison of the cost at different time slots under the different attack sequences assuming the controller implements perfect mitigation $\deltav_d[t] = \av[t], \ \forall t.$ It can be seen that the cost is greatest for MDP-based attacks, which validates its optimality. To investigate the attack impact under a practical mitigation strategy, we use a similar approach as described above and set  $\deltav_d[t] = \av[t]+\bv[t],$ where we generate $\bv[t]$ as a Gaussian distributed random variable with a standard deviation of $15$ units. From Fig.~\ref{fig:cost_imperfect}, it can be observed that even under the practical mitigation, the cost is greatest for the MDP-based attack sequence. Comparing Fig.~\ref{fig:general_model_cost} and Fig.~\ref{fig:cost_imperfect}, we also observe that the attack impact is greater for the practical mitigation compared with that of perfect mitigation (since perfect mitigation completely nullifies the attack's impact when it is detected).


\subsection{Quantifying the Cost of False Positives and Misdetections}
\label{sec:quant_MD_FP}
\begin{figure*}[!t]
\centering
\begin{subfigure}{0.35\textwidth}
\includegraphics[width=1\textwidth]{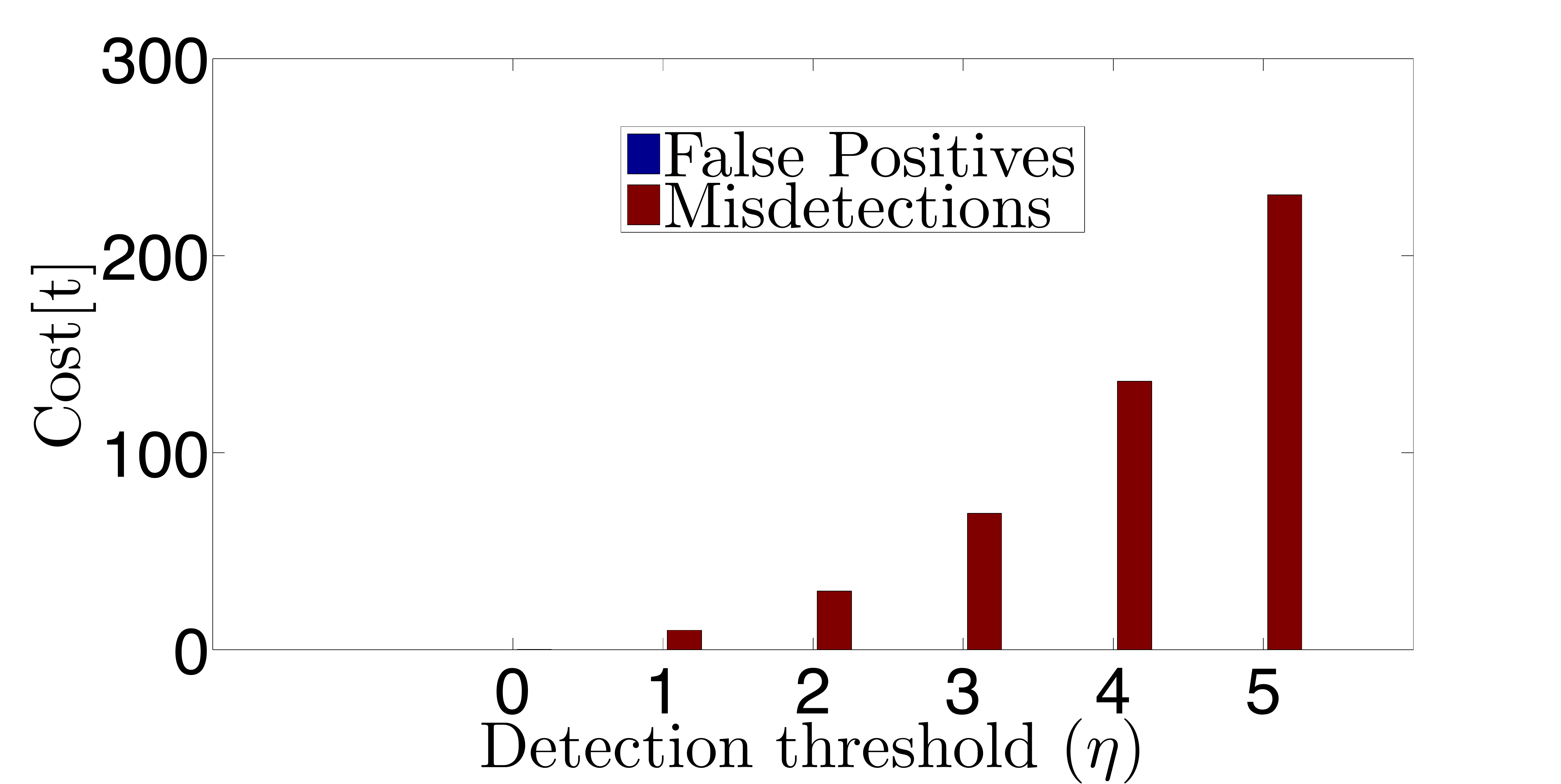}
\caption{$\sigma_{\text{mit}} = 0.$}
\end{subfigure}
~
\begin{subfigure}{0.35\textwidth}
\includegraphics[width=1\textwidth]{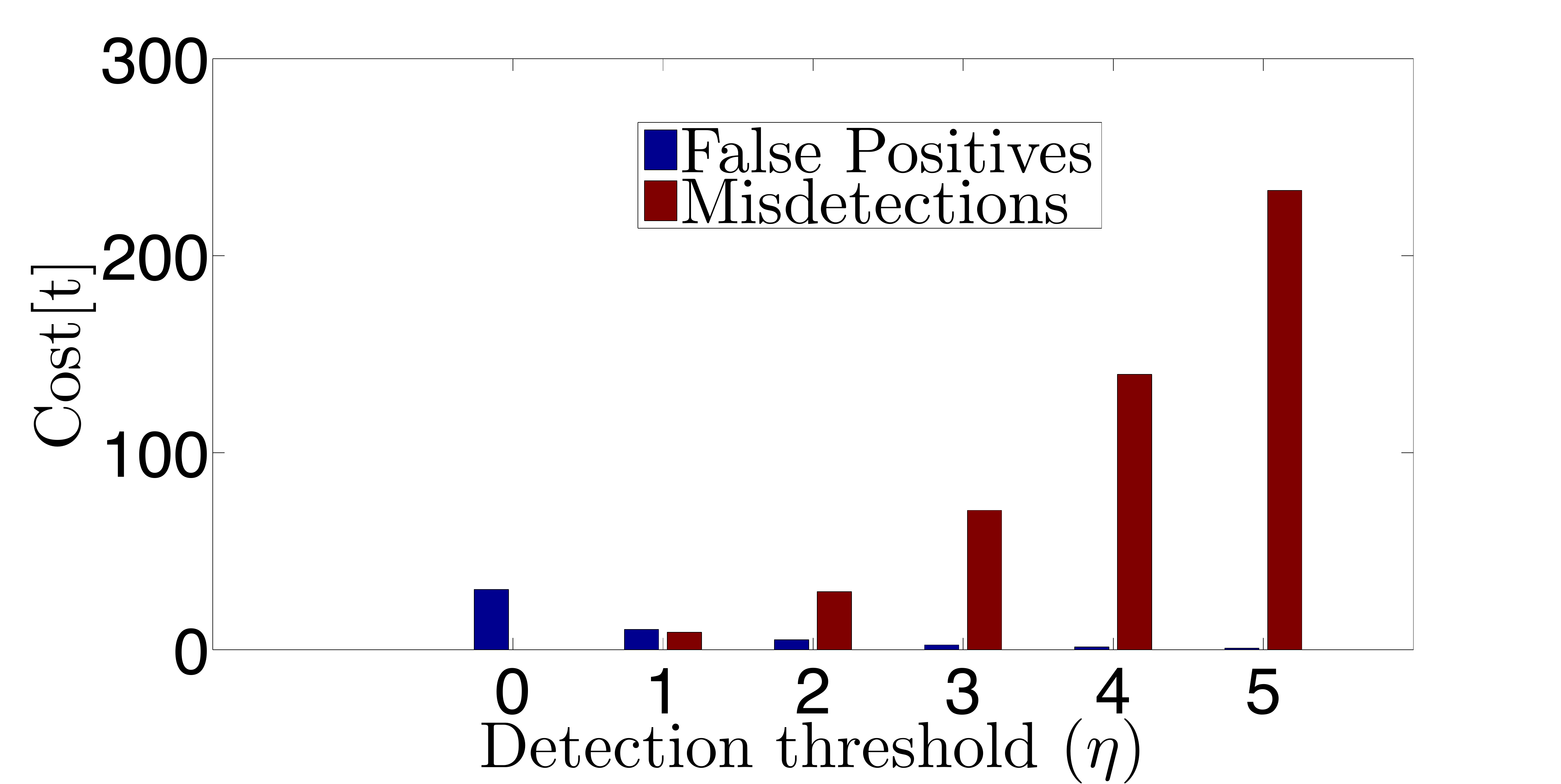}
\caption{$\sigma_{\text{mit}} = 5.$}
\end{subfigure} \\
~
\begin{subfigure}{0.35\textwidth}
\includegraphics[width=1\textwidth]{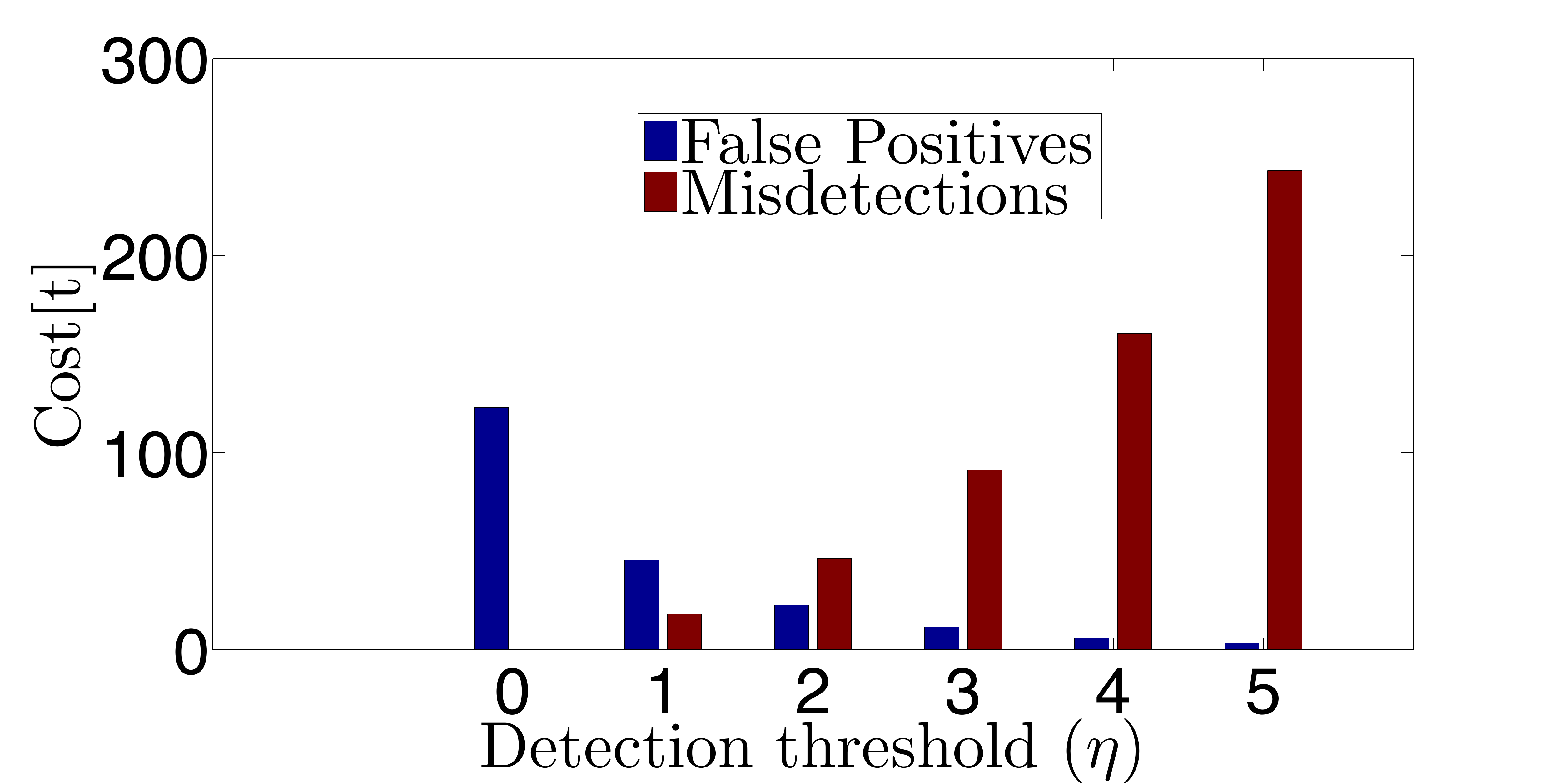}
\caption{$\sigma_{\text{mit}} = 10.$}
\end{subfigure}
~
\begin{subfigure}{0.35\textwidth}
\includegraphics[width=1\textwidth]{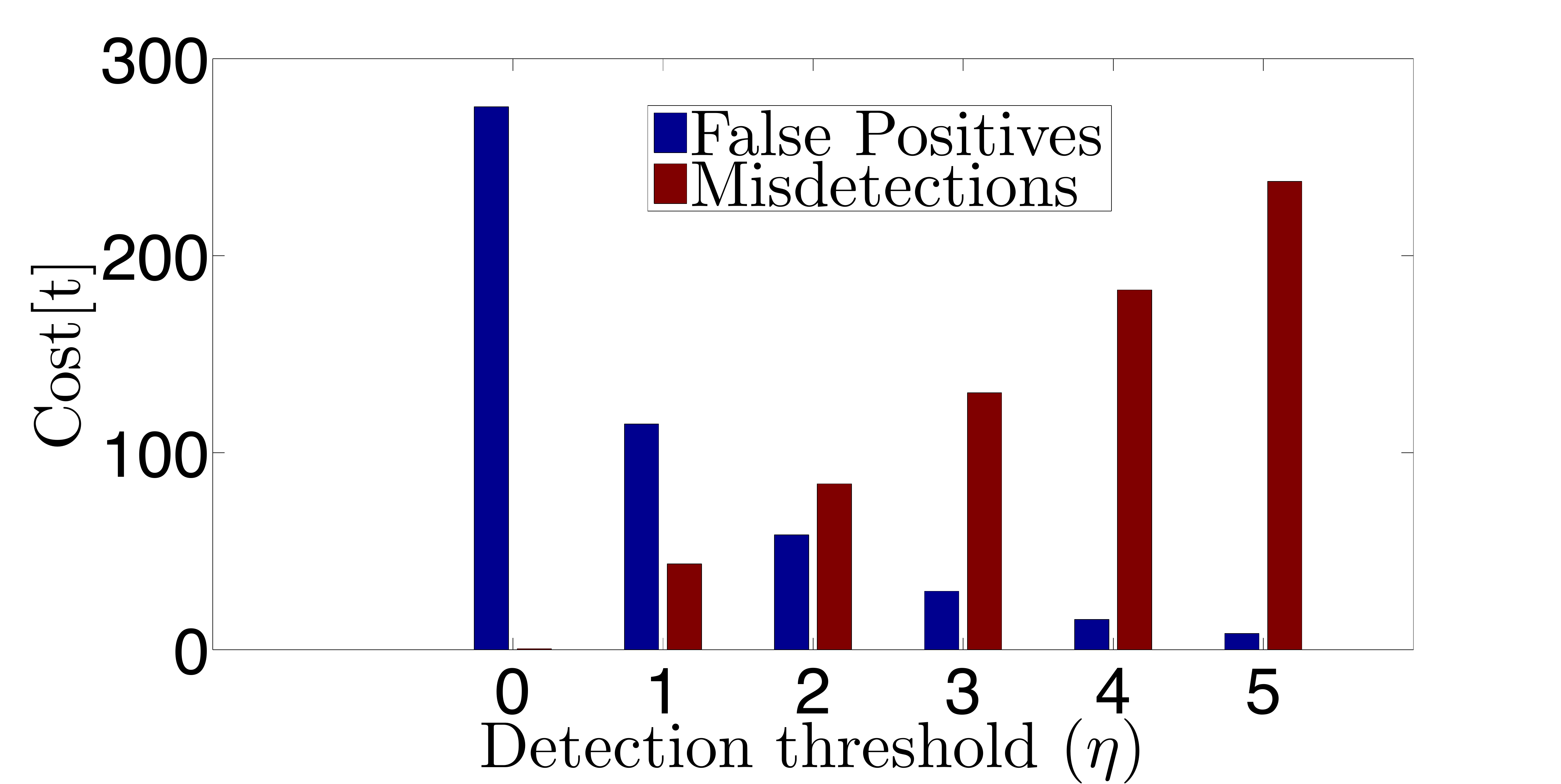}
\caption{$\sigma_{\text{mit}} = 15.$}
\end{subfigure}
\caption{Cost of FPs and MDs for different attack detection thresholds and standard deviation of the attack mitigation signal.}
\label{fig:Thold}
\end{figure*}

Next, we present simulation results to quantify the cost of FPs and MDs following the approach in Section~\ref{sec:MD_FP}. In our simulations, we consider the aforementioned practical attack mitigation. Fig.~\ref{fig:Thold} shows the cost of FPs and MDs for different detection thresholds $\eta$ and standard deviations of the attack mitigation signal $\sigma_{\text{mit}}$. We note that a low value of $\eta$ represents an aggressive detector, where as a high value of $\eta$ represents a conservative detector. For the mitigation signal, a low value of $\sigma_{\text{mit}}$ represents accurate mitigation, where as a high value represents inaccurate mitigation. In particular, $\sigma_{\text{mit}} = 0$ corresponds to perfect mitigation.

From these plots, we observe that as the attack detection threshold $\eta$ is increased, the cost of FP decreases, while the cost of MD increases. This result is intuitive -- a low detection threshold detects most attacks but also leads to a high number of FPs. Thus, the wrongly triggered mitigations will result in a high FP cost. On the other hand, a high detection threshold yields a low number of FPs, but also increases the number of MDs. The figures show a basic tradeoff between FPs and MDs, quantified in terms of the cost function. 

We also observe that these costs depend on the accuracy of the attack mitigation signal. E.g., when the accuracy is high (e.g.,  $\sigma_{\text{mit}} = 0, 5),$ the cost of FP is very low,
even for a low detection threshold. Thus, in this scenario, the system operator can choose a low detection threshold and obtain good system performance overall. However, when the accuracy of the mitigation signal is low (e.g., $\sigma_{\text{mit}} = 15),$ the cost of FP is very high for a low detection threshold. E.g., in Fig.~\ref{fig:Thold}(d), the cost of FP for $\eta = 0$ is greater than the cost of MD for $\eta = 5.$ In this scenario, the system operator must choose a high detection threshold to obtain an acceptable level of system performance. 
Thus, our result helps the system operator select an appropriate threshold that 
balances between the costs of FP and MD, depending on the accuracy of the mitigation signal.

Lastly, we note that for $\sigma_{\text{mit}} = 0$ (perfect mitigation), the cost of 
FP is zero for all detection thresholds. Under perfect mitigation, even 
if an FP event occurs, the controller can accurately estimate that the attack magnitude is zero (i.e., no attack). Thus, in this specific case, wrongly triggered mitigations do not increase the cost of FP.
We also note that for $\eta = 0,$ there are no MDs. Hence, the cost of MD in this case is nearly zero.

\subsection{Simulations for Voltage Control System}
Next, we perform simulations on the voltage control system using PowerWorld, which is a high fidelity power system simulator widely used in the industry \cite{PowerWorld}. 
All the simulations are performed on the IEEE 9-bus system shown in Fig.~\ref{fig:9_bus}, in which buses $1$, $2$, and $3$ are the generator buses, whereas buses $5$, $7$, and $9$ are the pilot buses.
The control matrix $\Bm$ is estimated using linear regression on the data traces of $\xv[t+1]-\xv[t]$ and $\uv[t]$ obtained in a PowerWorld simulation. We present the simulation results next.

\begin{figure}[!t]
\centering
\includegraphics[width=0.45\textwidth,trim={0 7cm 0 0}]{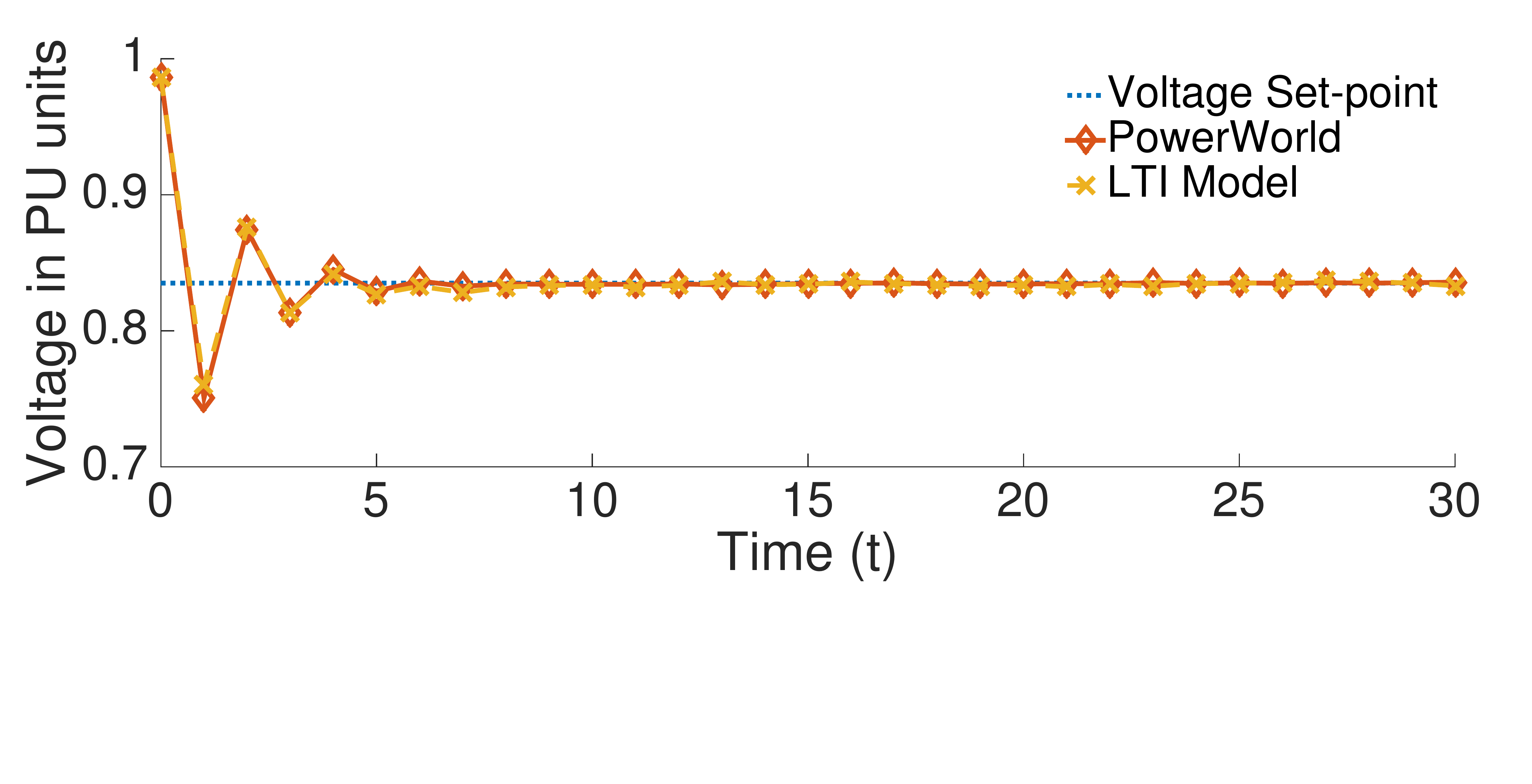}
\caption{Comparison between PowerWorld and LTI model.}
\label{fig:comaprsion_PW_LTI}
\vspace*{-0.4 cm}
\end{figure}

First, we verify the accuracy of the LTI model in approximating the real-world voltage control system by examining the voltage at pilot bus $5$. In our simulations, the voltage controller aims to adjust the voltage of this bus from an initial voltage of $1$~pu to a setpoint ($\xv_0$) of $0.835$~pu (base voltage of $230$~kV) by applying the control described in \eqref{eqn:control_vg}. 
Fig.~\ref{fig:comaprsion_PW_LTI} plots the bus voltage from $t = 1$ to $t =30$ obtained from the PowerWorld simulations, as well as the voltage values obtained from the LTI model. To average the effect of random measurement noise, we repeat the experiment $100$ times, and take the mean value. The two curves match
well in this figure, thus verifying the accuracy of the proposed LTI model.

\begin{figure}[!t]
\centering
\includegraphics[width=0.45\textwidth,trim={0 5.5cm 0 0}]{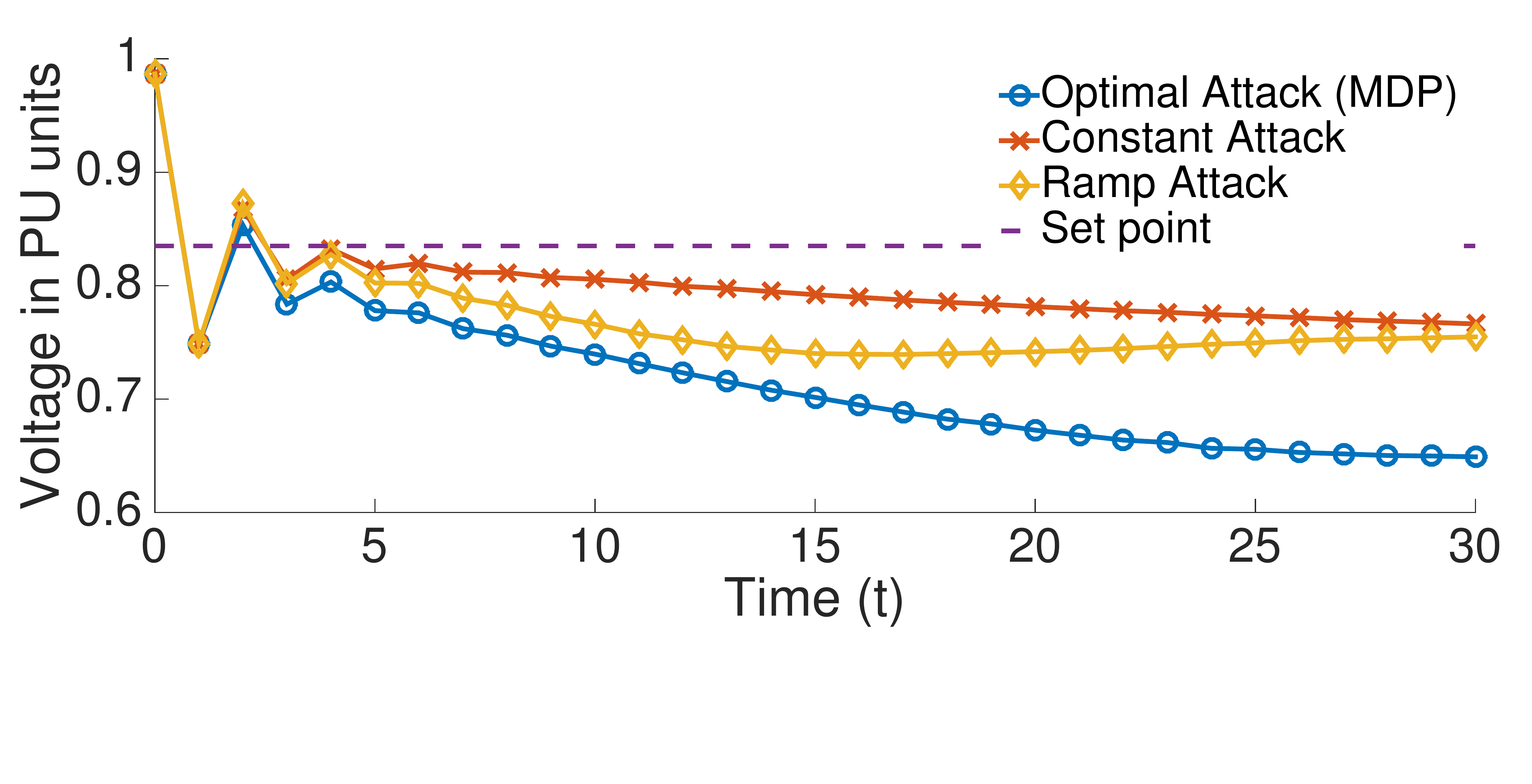}
\caption{Pilot bus voltage (Bus 5) under different attack sequences.}
\label{fig:voltage_state}
\vspace{-.5cm}
\end{figure}

Next, we simulate the impact of the proposed attacks on the voltage control system. 
We assume that the attacker has access to the voltage sensor of bus $5,$ and injects false measurements to mislead the controller. We compute the optimal attack sequence 
based on the LTI model using the MDP method implemented in MATLAB. 
To evaluate the attack impact, we run Monte Carlo simulations using the PowerWorld simulator by injecting the derived optimal attack into the voltage measurements, and implementing the control in \eqref{eqn:control_vg} based on the corresponding state estimate.
Fig.~\ref{fig:voltage_state} shows the pilot bus voltage (bus 5) for different attack sequences
with $\eta = 5$ 's and perfect attack mitigation. It can be observed that the pilot bus voltage deviates from the setpoint of $0.835$~pu, and the largest voltage deviation is seen under the optimal attack. In particular, over an attack duration of $30$ time slots, we observe that bus $5$ voltage deviates to $0.65$~pu under the optimal attack, a difference of about $0.2$~pu from its setpoint.

\begin{figure}[!t]
\centering
\includegraphics[width=0.45\textwidth,trim={0 7.5cm 0 0}]{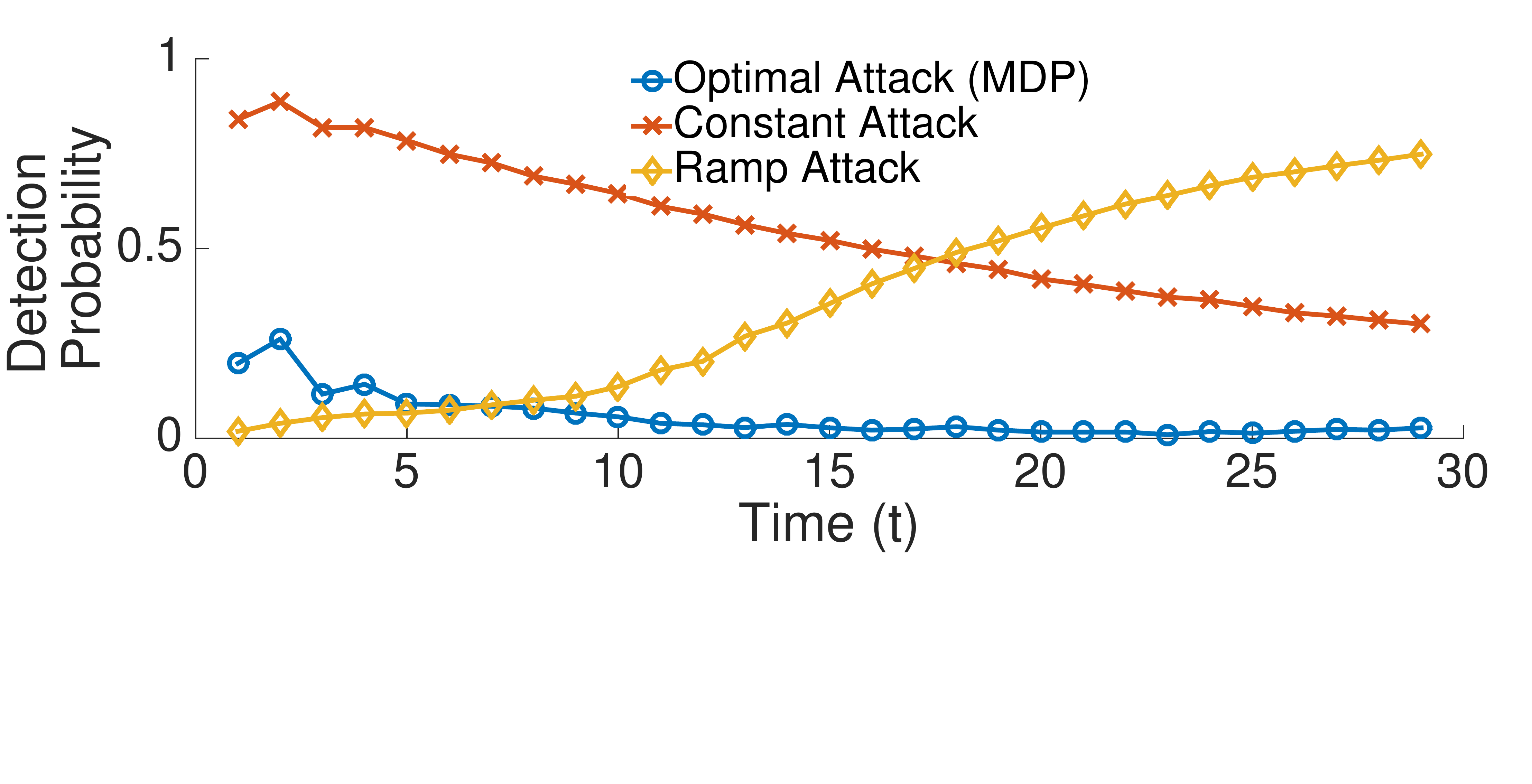}
\caption{Attack Detection probability for different attacks.}
\label{fig:voltage_probability}
\vspace{-.3cm}
\end{figure}

\begin{figure}[!t]
\centering
\begin{subfigure}{0.23\textwidth}
\includegraphics[width=1\textwidth]{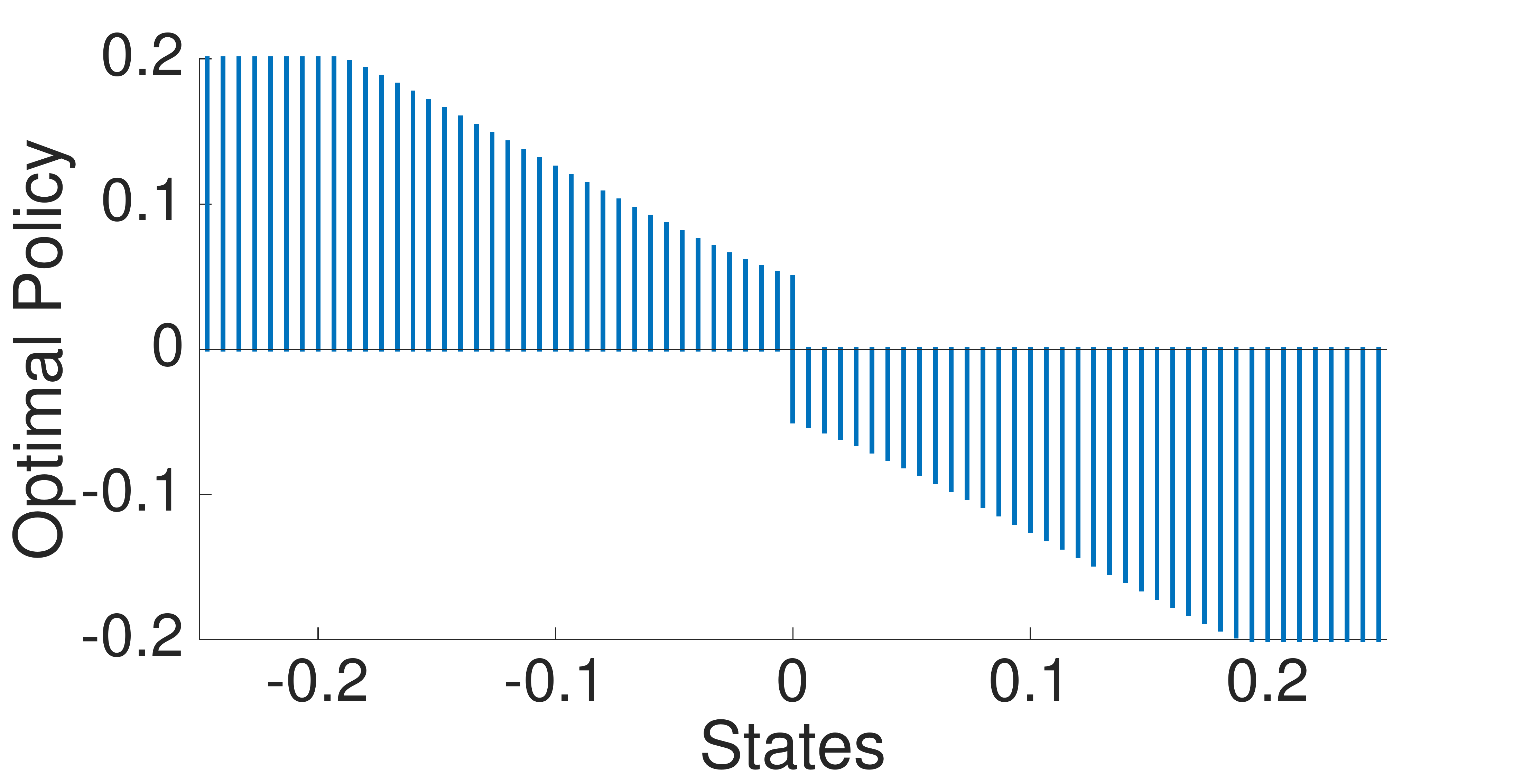}
\caption{}
\end{subfigure}
~
\begin{subfigure}{0.23\textwidth}
\includegraphics[width=1\textwidth]{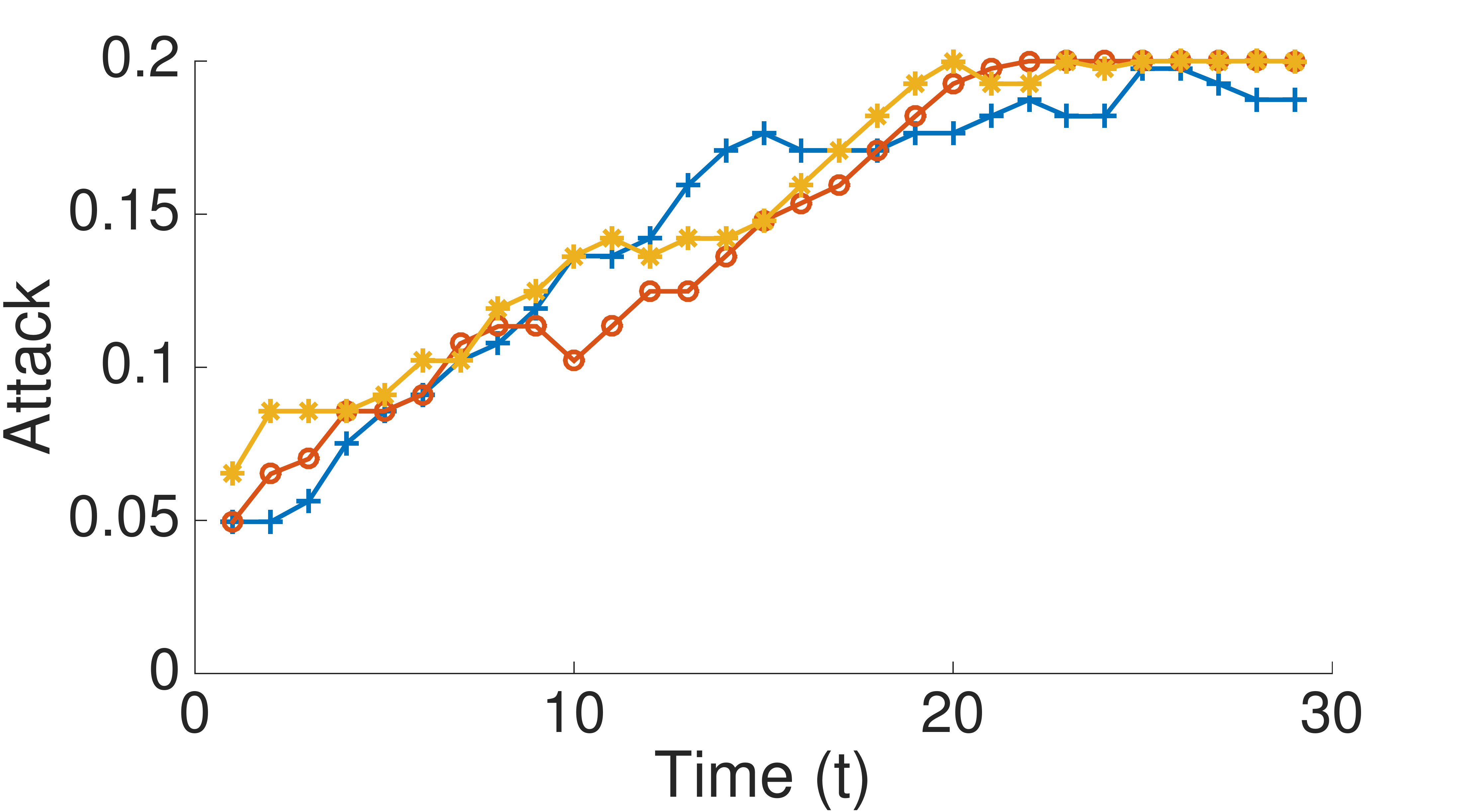}
\caption{}
\end{subfigure}
\caption{(a) Optimal policy for different system states as computed by the value iteration algorithm. (b) Optimal attack sequence for 3 Monte Carlo simulation instantiations.}
\label{fig:optimal_policy}
\end{figure}

Fig.~\ref{fig:voltage_probability} shows the attack detection probability under these attacks at different time instants. We also plot the optimal policy computed by the value iteration algorithm (Algorithm~1) in Fig.~\ref{fig:optimal_policy}a, and the optimal attack sequence for three Mote-Carlo instantiations in Fig.~\ref{fig:optimal_policy}b. We observe that the attack detection probability for a naive attack sequence (such as the ramp attack) increases with time, which results in nullifying its impact (due to attack mitigation). However, the optimal attack is crafted in a way such that the detection probability decreases over time. Consequently, the optimal attack causes a significant 
deviation of the pilot bus voltage from its setpoint.

\section{Conclusion}
\label{sec:Conclusion}
In this paper, we studied the performance of a CPCS with attack detection and reactive 
attack mitigation. We derived the optimal attack sequence that maximizes the state estimation error over the attack
time horizon using an MDP framework. Our results show that an arbitrarily constructed attack sequence will have little impact on the system since it will be detected, hence mitigated. The optimal attack sequence must be crafted to strike a balance between the stealthiness and the attack magnitude. Our results are useful for the system operator to assess the limit of attack impact and compare different attack detection and mitigation strategies. We also quantified the impact of FP and MD on the state estimation error, which 
helps select the right attack detection threshold depending on the accuracy of the attack mitigation signal.
We demonstrated the application of our results to the voltage control in a power system.

\bibliographystyle{ACM-Reference-Format}
\bibliography{bibliography}

\section*{Appendix A: Proofs of Lemma~5.1 and Corollary~5.2 }

\subsubsection*{Attack Detection Probability}
We start with the attack detection probability. To this end, we derive the relationship between the residual $\rv[t+1]$ and the KF estimation error $\ev[t].$ Using \eqref{eqn:process}, \eqref{eqn:Obs} and \eqref{eqn:Obs_Attack} in \eqref{eqn:res_defn}, followed by some algebraic manipulations, 
we obtain
\begin{align}
\rv[t+1] = \Cm \Am \ev[t] + \Cm \wv[t] + \av[t+1] + \vv[t+1]. \label{eqn:Ref12}
\end{align} 
Let $\rv_c[t+1]$ denote the conditional random variable given by
\begin{align}
\rv_c[t+1] & \defines  \rv[t+1] \big{|} \ev[t] = \ev ,\av[t+1] = \av \nonumber \\
& = \Cm \wv[t]  + \vv[t+1] +  \Cm \Am \ev + \av, \label{eqn:res_evol}
\end{align} 
where \eqref{eqn:res_evol} is obtained from \eqref{eqn:Ref12}.
According to the chi-square detection rule, the attack detection probability 
for a given value of $\ev[t] = \ev$ and $\av[t+1] = \av$ can be computed as
\begin{align}
\mathbb{P} (\rv_c[t+1]^T \Pm^{-1}_r \rv_c[t+1] \geq \eta). \label{eqn:attack_det}
\end{align}
From \eqref{eqn:res_evol}, it follows that the random variable $\rv_c[t+1]^T \Pm^{-1}_r \rv_c[t+1]$ follows a generalized chi-square distribution, which can be used to compute the probability 
in \eqref{eqn:attack_det}.

In particular for $n = m = 1,$ it follows that the attack is detected if
$\rv_c[t+1] \geq \sqrt{\eta \Pm_r},$ which is satisfied if (from \eqref{eqn:res_evol})
\begin{align}
& \Cm \wv[t] + \vv[t+1] \in  (-\infty, -\sqrt{\eta \Pm_r} - \Cm\Am\ev - \av] \nonumber \\ & \qquad \qquad \qquad \qquad \cup [\sqrt{\eta \Pm_r} - \Cm\Am\ev - \av, -\infty). \label{eqn:Ref3}
\end{align}
Alternately, the chi-square detector misses the attack if the noise terms satisfy
\begin{align}
&\Cm \wv[t] + \vv[t+1] \in \nonumber \\ 
& \qquad  \LSB -\sqrt{\eta \Pm_r}  - \Cm\Am\ev - \av , \sqrt{\eta \Pm_r} - \Cm\Am\ev - \av \RSB. \label{eqn:Ref1}
\end{align}
The probabilities of the events in \eqref{eqn:Ref3} and \eqref{eqn:Ref1} correspond
to attack detection and misdetection probabilities, which can be computed using the c.d.f. of the Gaussian distribution. 

\subsubsection*{MDP State Transition Probabilities}
Next, we compute the quantity $\mathbb{P} ( \ev_{\text{lb}} \leq \ev[t+1]  \leq \ev_{\text{ub}} \big{|} \ev[t] = \ev,\av[t+1] = \av).$ 
Recall the KF error evolution in \eqref{eqn:error_evol}.
Depending on the attack detection result $i[t+1],$ there are two cases:
\begin{itemize}
\item{Case~1:} When  $i[t+1] = 0,$ and $\ev_{\text{lb}} \leq \ev[t+1]  \leq \ev_{\text{ub}}.$
\item{Case~2:} When  $i[t+1] = 1,$ and $\ev_{\text{lb}} \leq \ev[t+1]  \leq \ev_{\text{ub}}.$
\end{itemize}
The quantity $\mathbb{P} ( \ev_{\text{lb}} \leq \ev[t+1]  \leq \ev_{\text{ub}} \big{|} \ev[t] = \ev,\av[t+1] = \av)$  can be computed as the sum of probabilities of the two cases. 
We investigate each case separately and derive its probability. 

\begin{itemize}
\item{Case~1:} 
Substituting $i[t+1] = 0,$ in \eqref{eqn:error_evol}, we obtain 
\begin{align}
\ev[t+1]  =  \Am_K \ev[t] + \Wm_K\wv [t]  -  \Km \av[t]   - \Km \vv[t+1]. \label{eqn:err_nodet}
\end{align}
Given $\ev[t] = \ev,$ and $\av[t+1] = \av,$ to have $\ev_{\text{lb}} \leq \ev[t+1]  \leq \ev_{\text{ub}},$ the noise terms must satisfy (from \eqref{eqn:err_nodet})
\begin{align}
& \Wm_K \wv[t] - \Km \vv[t+1] \in \nonumber \\ & \qquad \LSB \ev_{\text{lb}} - \Am_K\ev + \Km\av,   \ev_{\text{ub}}- \Am_K \ev + \Km \av \RSB. \label{eqn:Ref2}
\end{align}

In Case~1, the conditions \eqref{eqn:attack_det} and \eqref{eqn:Ref2} must be satisfied simultaneously, the probability of which can be computed as the joint probability of the two events given by the result
of Lemma~\ref{lem:trans_prob} (first expression of \eqref{eqn:TP_vector}).


In particular, for $n = m = 1,$ the conditions \eqref{eqn:Ref1} and \eqref{eqn:Ref2} must be satisfied simultaneously, the probability of which is given by
\begin{align}
\mathbb{P} \LB \begin{bmatrix}
-\sqrt{\eta \Pm_r} - \yv_1 \\
\ev_{\text{lb}} - \yv_2
\end{bmatrix} \ \leq \Xm \leq 
\begin{bmatrix}
\sqrt{\eta \Pm_r} - \yv_1 \\
\ev_{\text{ub}} - \yv_2
\end{bmatrix} \RB \label{eqn:Ref5}
\end{align}
where $\Xm \in \RR^{2n}$ is the concatenated vector 
given by
$$\Xm =  \begin{bmatrix}
\Cm \wv[t] + \vv[t] \\
\Wm_K \wv[t] - \Km\vv[t] 
\end{bmatrix},$$ and $\yv_1 = \Cm\Am\ev + \av$ and $\yv_2 = \Am_K\ev - \Km\av.$
The probability in \eqref{eqn:Ref5} can be computed using the Gaussian distribution as follows.   
Since $\wv[t]$ and $\vv[t+1]$ are Gaussian, the terms  $\Cm \wv[t] + \vv[t] $ and $\Wm_K \wv[t] - \Km\vv[t] $
are jointly Gaussian distributed. It is straightforward to note the mean value of the concatenated vector, i.e., $$\mathbb{E}[\Xm] = \begin{bmatrix}
\mathbb{E}[ \Cm \wv[t] + \vv[t] ]\\
\mathbb{E}[\Wm_K \wv[t] - \Km\vv[t] ]
\end{bmatrix} = {\bf 0},$$ and
its  covariance matrix is given by
\begin{align}
\text{Cov} & (\Xm)    = \begin{bmatrix}
\Cm \Qm \Wm^T_K + \Rm  & \Cm \Qm \Wm^T_K  - \Rm \Km^T \\
\Wm^T_K \Qm \Cm  - \Km \Rm^T & \Wm_K \Qm \Wm^T_K+ \Km \Rm  \Km^T
\end{bmatrix}.
\end{align}
Following the above arguments, we can compute \eqref{eqn:Ref5} using the c.d.f.
of the Gaussian distribution.

\item{Case~2:} Substituting $i[t+1] = 1,$ in \eqref{eqn:error_evol}, we obtain 
\begin{align}
\ev  [t+1]  =  \Am_K \ev + \Wm_K\wv [t]  -  \Km (\av - \deltav)   - \Km \vv[t+1]. \label{eqn:err_det}
\end{align}
Given $\ev[t] = \ev,$ $\av[t+1] = \av,$ and $\deltav[t+1] = \deltav$ to have $\ev_{\text{lb}} \leq \ev[t+1]  \leq \ev_{\text{ub}},$ the noise terms must satisfy (from \eqref{eqn:err_det})
\begin{align}
& \Wm_K \wv[t] - \Km\vv[t+1] \in \nonumber \\ & \qquad  \big{[} \ev_{\text{lb}} - \Am_K\ev + \Km(\av - \deltav),   \ev_{\text{ub}}- \Am_K\ev + \Km(\av - \deltav) \big{]}. \label{eqn:Ref4}
\end{align}
In Case~2, the conditions \eqref{eqn:Ref3} and \eqref{eqn:Ref4} must be satisfied simultaneously, the probability of which can be computed as the joint probability of the two events, given by the result
of Lemma~\ref{lem:trans_prob} (second expression of \eqref{eqn:TP_vector}).

For $n = m = 1,$ the conditions \eqref{eqn:Ref3} and \eqref{eqn:Ref2} must be satisfied simultaneously, the probability of which is provided in the result of Corollary~\ref{cor:trans_prob} (second
and third expressions of \eqref{eqn:norm_cdf}). The probabilities can be computed using the
c.d.f. of the Gaussian distribution similar to Case~1.
 
\end{itemize}

\section*{Appnedix B: MDP Discretization}
In this appendix, we provide details of the MDP discretization procedure of Section~\ref{sec:MDP_Discrete}. Formally, we define the discretized MDP by a tuple $(\Xi,\mathcal{A},\overline{\mathcal{T}},\overline{R}).$ Here in, $\Xi$ denotes a discretized version of the original state space $\mathcal{E},$ given by $\Xi = \{ \xi_1,\dots,\xi_N \},$ 
where $N$ is the number descritization levels, and $\overline{\mathcal{T}} (\xi_i,\av,\xi_j)$, $\overline{R}(\xi_i,\av,\xi_j)$ and 
$\overline{V} (\xi_i)$ denote respectively the state transition probabilities, the reward, 
and the value function of the discretized MDP. Next, we elaborate the three steps 
involved in the MDP discretization as stated in Section~\ref{sec:MDP_Discrete}.

\begin{figure}
\centering
\includegraphics[width=0.45\textwidth]{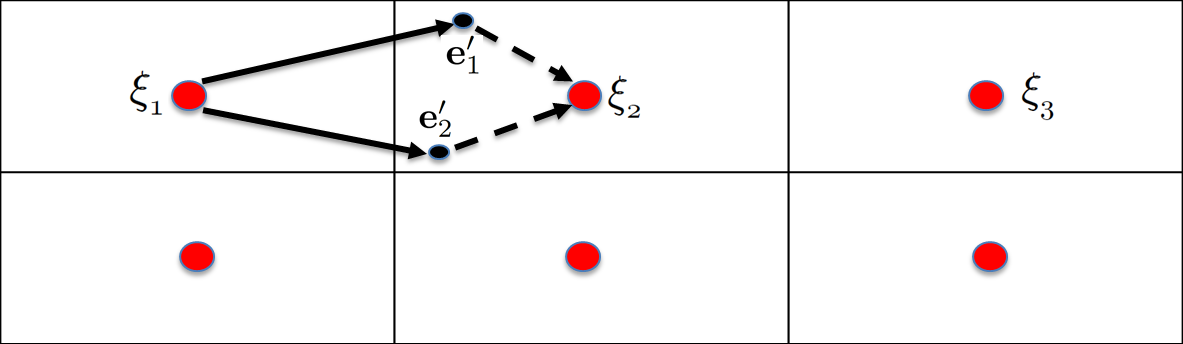}
\caption{A pictorial representation of the discretization procedure.}
\label{fig:grid}
\end{figure}

We start with Step~1, i.e., construction of the discretized MDP from the original continuous MDP. 
A pictorial illustration of the discretization procedure is shown in Fig.~\ref{fig:grid}.
In this figure, points $\xi_1,\xi_2,\dots$ represent a discretized version of the 
original MDP's continuous state space. Note that the points $\xi_1,\xi_2,\dots$ are a subset of the original state space. 
The arrows in Fig.~\ref{fig:grid} show a mapping between state transitions of
the continuous MDP to those of the discretized MDP. The mapping is based on the following logic. 
Consider all the state transitions in the continuous MDP from $\ev[t] = \xi_i, \ i = 1,2,\dots,$ to a state $\ev[t+1] = \ev^{\prime} \notin \{\xi_1, \xi_2,\dots \}$ under an action $\av.$
In the discretized MDP, all such transitions are mapped to a state $\xi_i, \ i = 1,2,\dots, $ 
that is nearest to $\ev^{\prime}.$
E.g., in Fig.~\ref{fig:grid}, all the state transitions in the continuous MDP from $\ev[t] = \xi_1$ to the states $\ev^{\prime}_1$ and $\ev^{\prime}_2$ are mapped to $\xi_2$ (since $\xi_2$ is closest to 
$\ev^{\prime}_1$ and $\ev^{\prime}_2$ in the discretized state space). 
The state transition probabilities from $\xi_1$ to $\xi_2$ in the discretized MDP is computed
as the sum (and in the limiting case, the integration) of all such state transition probabilities of the continuous MDP. 

Based on this logic, a mathematically rigorous way to compute $\overline{\mathcal{T}}(\xi_i,\av, \xi_j)$ from $\mathcal{T}(\xi_i,\av, \xi_j)$ is given by
\begin{align*}
\overline{\mathcal{T}}(\xi_i,\av, \xi_j) = \mathbb{P} (\xi_j \big{|} \xi_i,\av) = \int_{\ev^{\prime} \in B(\xi_j) }
\mathcal{T}( \xi_i,\av,\ev^{\prime}), 
\end{align*}
where $B(\xi_j)$ denotes the set of points $\ev^{\prime} \in \mathcal{E}$ which are closer to $\xi_j$ than any other point $\xi_k \in \Xi, k \neq j.$ Mathematically, $B(\xi_j)$ 
is given by
\begin{align*}
B(\xi_j) = \{ \ev^{\prime} \in \mathcal{E} \big{|} d(\ev^{\prime}, \xi_j) \leq d(\ev^{\prime}, \xi_k), 1 \leq k \leq N, k \neq j \}, 
\end{align*}
where $d(\xv,\yv)$ denotes the Euclidean distance between the points $\xv$ and $\yv,$ i.e. $d(\xv, \yv) = ||\xv- \yv||.$

The immediate expected reward in the discretized MDP $\overline{R}(\xi_i,\av, \xi_j)$
can be computed as $\overline{R}(\xi_i,\av, \xi_j) = \sum^N_{j = 1}\overline{\mathcal{T}}(\xi_i,\av, \xi_j) || \xi_j ||^2.$

Next, we proceed to Step~2 of the discrterization procedure, i.e., computing the optimal policy of the discretized MDP. We use the notations $\overline{\pi}^*$ to denote the optimal policy of the discretized MDP and $\overline{V}^{*}(\xi_i)$ to denote the optimal state value function of state $\xi_i \in \Xi$.
They can be computed using the value iteration algorithm specified as Algorithm~1.

Finally, we proceed to Step~3 of the discrterization procedure, i.e., mapping the optimal 
policy of the discretized MDP to a near-optimal policy of the continuous MDP. 
First note that the optimal policy of the discretized MDP computed in Algorithm~1 cannot be
directly applied to the continuous MDP, since we do not know the optimal policy 
for a state $\ev \in \mathcal{E}$ that is not in the discretized state space $\Xi.$
To address this issue, we use the nearest neighbour approximation, i.e.,
for a state $\ev \notin \Xi,$ we choose an action based on
the policy of its nearest neighbour,
$\pi(\ev) = \bar{\pi}^*(\xi_i), \ \text{where} \ \xi_i = \argmin_{1 \leq i \leq N} ||\ev-\xi_i||$.


\end{document}